\documentclass[aps, prd, amssymb, amsmath, amsfonts, eqsecnum, showpacs, nofootinbib, floatfix, twocolumn, twoside, a4paper, superscriptaddress]{revtex4-2}

\usepackage[T3,T1]{fontenc}
\usepackage{mathrsfs} % For \mathscr fonts
\usepackage{bm} % For \bm fonts
\makeatletter
\@ifclassloaded{beamer}
  { % if this is a beamer document, use sans-serif fonts
    \typeout{UsePackages: Detected beamer}
    \usepackage{tgheros}
    
  }
  { % otherwise:
    \typeout{UsePackages: Did not detect beamer}
    \ifx\asybeamer\undefined % use times for articles
    \typeout{UsePackages: Detected article}
    \usepackage[varg]{txfonts} % For math
    \usepackage{tgtermes} % For text
    \else % use beamer fonts for asy documents with beamer
    \typeout{Fonts: Detected asy for beamer}
    \usepackage{tgheros}
    
    \fi
  }
\usepackage{graphicx} %
\graphicspath{{./figures/}}
\usepackage[x11names,svgnames,rgb]{xcolor} %
\usepackage{xspace}
\usepackage{braket}
\usepackage{accents}
\usepackage{siunitx}
\usepackage{xfrac}
\usepackage{mathtools}
\DeclareSymbolFontAlphabet{\mathrm}{operators}

\definecolor{CiteColor}{rgb}{0.18039, 0.18824, 0.57255}
\definecolor{UrlColor} {rgb}{0.741, 0.173, 0.000}
\definecolor{DarkUrlColor} {rgb}{0.500, 0.110, 0.000}
\definecolor{LinkColor}{rgb}{0.25098, 0.47843, 0.04706}
\makeatletter %
\newcommand{\ShowFont}{%
  \typeout{The main font is \f@encoding \space \f@family \space %
    \f@series \space \f@shape \space at \f@size pt.}%
  \typeout{The math font sizes are \tf@size pt (main), \sf@size pt %
    (script), and \ssf@size pt (scriptscript).}%
  \typeout{The linewidth is \the\linewidth}} %
\makeatother %

\usepackage{xargs}

\usepackage{academicons}

\usepackage{dcolumn}% Align table columns on decimal point
\usepackage{longtable}
\usepackage{relsize}
\usepackage{bm}% bold math
\usepackage{tabularx}
\usepackage{booktabs}
\usepackage{xspace}
\usepackage[utf8]{inputenc}
\usepackage[normalem]{ulem}
\usepackage[colorlinks]{hyperref}

\usepackage{orcidlink}

\newcommand{\mpr}{m^{\prime}}
\newcommand{\ptimes}{\ensuremath{\mathbin{\substack{+\raisebox{0.1ex}{\vphantom{+}} \\[-0.3ex] \times}}}}
\newcommand{\fmeco}{f_{\rm{MECO}}}
\newcommand{\fisco}{f_{\rm{ISCO}}}
\newcommand{\fdamp}{f_{\rm{damp}}}
\newcommand{\frd}{f_{\rm{RD}}}

\newcommand{\dphi}{\delta\hat{\varphi}}
\newcommand{\dpi}{\delta\hat{p}}
\newcommand{\logB}[1]{\ensuremath{\log_{10}\mathcal{B}^{\mathrm{#1}}}\xspace}
\newcommand{\QGR}{\ensuremath{\mathcal{Q}_{\mathrm{GR}}}\xspace}

\newcommand{\phD}{\mbox{\textsc{IMRPhenomD}}\xspace}
\newcommand{\phP}{\mbox{\textsc{IMRPhenomP}}\xspace}
\newcommand{\phPvHM}{\mbox{\textsc{IMRPhenomPv}3\textsc{HM}}\xspace}
\newcommand{\phX}{\mbox{\textsc{IMRPhenomX}}\xspace}
\newcommand{\phXAS}{\mbox{\textsc{IMRPhenomXAS}}\xspace}
\newcommand{\phXP}{\mbox{\textsc{IMRPhenomXP}}\xspace}
\newcommand{\phXHM}{\mbox{\textsc{IMRPhenomXHM}}\xspace}
\newcommand{\phXPHM}{\mbox{\textsc{IMRPhenomXPHM}}\xspace}
\newcommand{\phXPHMMSA}{\mbox{\textsc{IMRPhenomXPHM-MSA}}\xspace}
\newcommand{\phXPHMST}{\mbox{\textsc{IMRPhenomXPHM-SpinTaylor}}\xspace}
\newcommand{\phXOfoura}{\mbox{\textsc{IMRPhenomXO}4a}\xspace}
\newcommand{\phXPNR}{\mbox{\textsc{IMRPhenomXPNR}}\xspace}
\newcommand{\phXPT}{\mbox{\textsc{IMRPhenomXP\_NRTidal}v2}\xspace}
\newcommand{\phXAST}{\mbox{\textsc{IMRPhenomXAS\_NRTidal}v2}\xspace}
\newcommand{\phXPTv}{\mbox{\textsc{IMRPhenomXP\_NRTidal}v3}}
\newcommand{\phXASTv}{\mbox{\textsc{IMRPhenomXAS\_NRTidal}v3}}

\newcommand{\NRTidalvThree}{\mbox{\textsc{NRTidal}v3}\xspace}
\newcommand{\NRTidalvTwo}{\mbox{\textsc{NRTidal}v2}\xspace}

\newcommand{\seobnrrom}{\textsc{SEOBNRv}4\textsc{\_ROM}\xspace}
\newcommand{\seobnrHMrom}{\textsc{SEOBNRv}4\textsc{HM}\textsc{\_ROM}\xspace}

\newcommand{\bilby}{\mbox{\textsc{Bilby}}\xspace}
\newcommand{\bilbytgr}{\mbox{\textsc{Bilby TGR}}\xspace}

\newcommand{\lalsuite}{\mbox{\textsc{LALSuite}}\xspace}

\newcommand{\lalsimulation}{\mbox{\textsc{LALSimulation}}\xspace}
\newcommand{\dynesty}{\mbox{\textsc{Dynesty}}\xspace}
\newcommand{\numpy}{\mbox{\textsc{NumPy}}\xspace}
\newcommand{\scipy}{\mbox{\textsc{SciPy}}\xspace}
\newcommand{\matplotlib}{\mbox{\textsc{Matplotlib}}\xspace}

\newcommand{\checkmarkcut}{
  \checkmark\makebox[2pt][r]{\raisebox{0.5ex}{\normalfont\symbol{'26}}}
}

\newcommand{\starcutcheckmark}{
  \checkmark\makebox[2pt][r]{\raisebox{0.5ex}{\normalfont\symbol{'26}}}\makebox[4pt][r]{\raisebox{0.5ex}{\normalfont{*}}}
}

\DeclareMathAlphabet{\mathcalstd}{OMS}{cmsy}{m}{n}
\DeclareMathAlphabet{\mathpzc}{OT1}{pzc}{m}{it}

\newcommand{\Nikhef}{Nikhef -- National Institute for Subatomic Physics, Science Park 105, 1098 XG Amsterdam, The Netherlands}

\newcommand{\UU}{Institute for Gravitational and Subatomic Physics (GRASP), \mbox{Utrecht University}, Princetonplein 1, 3584 CC Utrecht, The Netherlands}

\newcommand{\UCLouvain}{Centre for Cosmology, Particle Physics and Phenomenology - CP3, Universit\'{e} Catholique de Louvain, Louvain-La-Neuve, B-1348, Belgium}
\newcommand{\ROB}{Royal Observatory of Belgium, Avenue Circulaire, 3, 1180 Uccle, Belgium}

\newcommand{\bham}{School of Physics and Astronomy and Institute for Gravitational Wave Astronomy, University of Birmingham, Edgbaston, Birmingham, B15 2TT, United Kingdom}

\definecolor{dodgerblue}{HTML}{1E90FF}
\hypersetup{citecolor=dodgerblue}

\definecolor{RED}{HTML}{F5054F}
\definecolor{LIGHT_ORANGE}{HTML}{FDAA48}
\definecolor{DARK_ORANGE}{HTML}{FF5B00}
\definecolor{LIGHT_BLUE}{HTML}{448EE4}
\definecolor{DARK_BLUE}{HTML}{0343df}
\definecolor{LIGHT_GREEN}{HTML}{40C53C}
\definecolor{DARK_GREEN}{HTML}{02590F}

\definecolor{lapislazuli}{rgb}{0.15, 0.38, 0.61}

\newcommand{\Hz}{\ensuremath{\,\mathrm{Hz}}\xspace}

\newcommand{\flow}{\ensuremath{f_{\mathrm{low}}}\xspace}

\newcommand{\fhigh}{\ensuremath{f_{\mathrm{high}}}\xspace}

\newcommand\abs[1]{\ensuremath{\lvert#1\rvert}}

\newcommand\inp[2]{\langle #1 \,|\, #2 \rangle}

\def \msun  {\rm{M}_\odot}
\newcolumntype{d}{D{.}{.}{-1}}
\newcolumntype{C}{>{\centering}X}

\begin{document}

%\title{An improved parametrized test of general relativity with state-of-the-art IMRPhenomX waveform models}
%\title{An improved parametrized test of general relativity with IMRPhenomX models: including higher-harmonics and precession}
\title{An improved parametrized test of general relativity using the \phX waveform family: Including higher harmonics and precession}

\author{Soumen Roy\orcidlink{0000-0003-2147-5411}}
\email{soumen.roy@uclouvain.be}
\affiliation{\UCLouvain}
\affiliation{\ROB}
\affiliation{\Nikhef}
\affiliation{\UU}

\author{Maria Haney\orcidlink{0000-0001-7554-3665}}
\affiliation{\Nikhef}

\author{Geraint Pratten~\orcidlink{0000-0003-4984-0775}}
\affiliation{\bham}

\author{Peter~T.~H.~Pang~\orcidlink{0000-0001-7041-3239}}
\affiliation{\Nikhef}
\affiliation{\UU}

\author{Chris Van Den Broeck}
\affiliation{\Nikhef}
\affiliation{\UU}

\begin{abstract}

When testing general relativity (GR) with gravitational wave observations, parametrized tests of deviations from the expected strong-field source dynamics are one of the most widely used techniques. We present an updated version of the parametrized framework with the state-of-art \phX waveform family. Our new framework incorporates deviations in the dominant mode as well as in the higher-order modes of the waveform. We demonstrate that the missing physics of either higher-order modes or precession in the parametrized model can lead to a biased conclusion of false deviation from GR. Our new implementation mitigates this issue and enables us to perform the tests for highly asymmetric and precessing binaries without being subject to systematic biases due to missing physics. Finally, we apply the improved test to analyze events observed during  the second half of the third observing run of LIGO and Virgo (O3b). We provide constraints on GR deviations by combining O3b results with those from previous observation runs.  Our findings show no evidence for violations of GR.

\end{abstract}
\maketitle

\section{Introduction}

The detections of gravitational waves (GWs) from compact binary coalescences by Advanced LIGO~\cite{LIGOScientific:2014pky} and Advanced Virgo~\cite{VIRGO:2014yos} have opened up a new laboratory for testing general relativity (GR). This includes exploring the two-body dynamics in strong gravitational fields with velocities approaching the speed of light, the nature of compact objects, the properties of GW propagation, and the presence of extra dimensions. In this context, one of the most popular techniques are parametrized tests of deviations from the general-relativistic source dynamics of compact binary mergers.
%is the parametrized test of GR using GWs.

%GR is the simplest and most successful theory of gravity to date, and it has been rigorously validated by various experimental tests in our Solar System~\cite{Will:2014kxa}, observations of binary pulsars~\cite{Wex:2014nva, Shao:2017gwu}, cosmological data~\cite{Ferreira:2019xrr}, GWs~\cite{LIGOScientific:2016lio, LIGOScientific:2016dsl, LIGOScientific:2018dkp, LIGOScientific:2019fpa, LIGOScientific:2020tif, LIGOScientific:2021sio}, and massive BHs~\cite{Genzel:2024vou}. Recently, the evidence of a stochastic astrophysical GW background with pulsar timing arrays~\cite{2023arXiv231007537B, 2023arXiv230711665C} has also led to new gravitational tests. Despite its overall success, we do not know how to reconcile GR with quantum mechanics and how (or if) to employ GR to explain certain cosmological phenomena, such as dark matter and dark energy. These limitations have motivated the development of alternative theories of gravity , such as the Brans-Dicke~\cite{Brans:1961sx}, Horndeski scalar-tensor~\cite{Horndeski:1974wa}, Aether~\cite{Jacobson:2004ts}, Einstein–Gauss-Bonnet~\cite{Nojiri:2005vv, DeFelice:2010aj}, Chern-Simons~\cite{Jackiw:2003pm, Alexander:2009tp} theories, and the effective-field-theory extension of GR~\cite{Endlich:2017tqa}, all of which have survived a wide range of experimental tests~\cite{Will:2014kxa, Yunes:2016jcc, Lyu:2022gdr, Takeda:2023wqn, Silva:2022srr}.

The GW signals from inspiralling compact binaries are modeled using post-Newtonian (PN) expansion, expressed as a series expansion in the orbital velocity. The post-inspiral part of the waveform is modeled using a set of phenomenological coefficients, which are determined by calibrating with numerical relativity and black hole perturbation theory for the ringdown part. Developing an accurate GW waveform model that describes the entire evolution of compact binaries is essential for determining source properties, validating the predictions of GR, and exploring the potential alternative theories of gravity. However, the development of waveforms within specific alternative theories has not yet reached a level of maturity sufficient to perform model comparisons with GR. While numerical-relativity simulations in beyond-GR theories are becoming more prevalent~\cite{East:2020hgw, Bezares:2021dma, Figueras:2021abd, AresteSalo:2022hua, AresteSalo:2023mmd, AresteSalo:2023hcp, Corman:2022xqg, Corman:2024vlk, Okounkova:2019dfo, Okounkova:2020rqw, Cayuso:2023aht}, there are not yet sufficient simulations in one specific theory to calibrate a semi-analytic inspiral-merger-ringdown model in that theory. Moreover, the true theory of gravity may be an alternative to GR that we are not yet theoretically aware of. This has led to the development of theory-agnostic approaches, which are classified into two categories: either verifying the consistency between the prediction of GR and data or introducing a deviation parameter to quantify the degree to which the GR prediction agrees with observed data.

Alternative theories of gravity often suggest deviation in the numerical values of the coefficients compared to those derived from Einstein's GR. This has prompted the development of theory-agnostic inspiral parametrized tests of GR~\cite{Blanchet:1994ez,Arun:2006yw, Arun:2006hn, Yunes:2009ke, Mishra:2010tp, Agathos:2013upa, Li:2011cg, Meidam:2017dgf}, which look for the departure from GR by introducing parametric deformations to the PN phasing coefficients. The same framework can be used in search pipelines to detect non-GR signals~\cite{Narola:2022aob, Sharma:2023djw}. In the inspiral parametrized test, we introduce a deviation parameter (also referred to as the non-GR or beyond-GR parameter) to the PN phasing coefficients, allowing for a generic departure from GR. A similar approach can be applied to the post-inspiral coefficients defined in phenomenological inspiral–merger–ringdown (IMR) waveform models~\cite{Meidam:2017dgf}. For each coefficient, we measure that deviation parameter by performing Bayesian parameter estimation analysis. If the resulting posterior of the deviation parameter is consistent with zero, we then provide a quantitative indication of the degree to which GR describes the data. In contrast, finding inconsistency with zero may indicate that GR does not agree with the observed data, which will lay the foundation for a new theoretical model. However, claiming a violation of GR would require extensive studies to rule out false identifications of GR deviations. Potential sources of false deviations include noise systematics, waveform systematics (missing physics in the source model such as spin precession and higher-order modes~\cite{Pang:2018hjb, Maggio:2022hre, Chandramouli:2024vhw}, eccentricity~\cite{Narayan:2023vhm, Shaikh:2024wyn}), astrophysical environmental effects~\cite{CanevaSantoro:2023aol, Roy:2024rhe}, and gravitational lensing~\cite{Mishra:2023vzo, Wright:2024mco, Liu:2024xxn}, among others. A comprehensive list of these potential sources is provided in Ref.~\cite{2024arXiv240502197G}.

Parametrized tests of the phasing coefficients within a Bayesian framework have been extensively developed for the LIGO-Virgo-KAGRA (LVK) analyses of testing GR with GW transient catalogues~\cite{LIGOScientific:2019fpa, LIGOScientific:2020tif, LIGOScientific:2021sio}. The results have been reported for two implementations: the Test Infrastructure for General Relativity (TIGER) approach~\cite{Li:2011cg, Agathos:2013upa, Meidam:2017dgf}, which is based on a frequency-domain phenomenological waveform family and the Flexible Theory-Independent (FTI) approach~\cite{LIGOScientific:2018dkp, Mehta:2022pcn}, which can be applied to any aligned-spin frequency-domain waveform model.

In this work, we develop a TIGER framework for the parametrized tests using the \phX waveform family with high-order modes and precession. We incorporate the deviation parameter into the phasing coefficients of the dominant quadrupole waveform \phXAS model. In the inspiral phase, we introduce a fractional deviation to each PN coefficient, starting from  0PN up to 3.5PN order. We also introduce absolute deviation parameters at -1PN and 0.5PN, as many alternative theories predict nonzero value at those PN orders, but they are uniquely zero in Einstein's theory. In the post-inspiral part, we introduce the fractional deviation to the phenomenological coefficients of the model itself. During the inspiral phase, the phasing coefficients in the higher-order modes are connected through a scaling relation, which enables us to propagate the same deviation into the high-order modes of the waveform. 

The earlier version of TIGER was developed by modifying the phasing coefficients of the \phD approximant~\cite{Meidam:2017dgf, Husa:2015iqa, Khan:2015jqa}. Since \phD forms the baseline of \phPvHM~\cite{Khan:2019kot}, this allowed the corrections to propagate into the higher-order modes. However, \phPvHM is not well suited for highly asymmetric or strongly precessing binaries, and it becomes comparatively slow when precession and higher-order modes are enabled. In contrast, the new TIGER framework employs the more accurate and computationally efficient \phXPHM waveform model~\cite{Pratten:2020ceb}. The \phX family was calibrated to a substantially larger set of numerical-relativity simulations~\cite{Pratten:2020ceb, Pratten:2020fqn, Pratten:2020ceb, Garcia-Quiros:2020qpx}, improving accuracy across a broader region of the parameter space. The \phX family also incorporates several optimizations and an efficient twisting-up procedure, making \phXPHM significantly faster to evaluate than \phPvHM and therefore better suited for the large-scale Bayesian analyses required for parameterized tests with TIGER. Our new implementation thus enables computationally efficient and more accurate tests of GR for highly asymmetric and precessing binaries, without incurring systematic biases from missing physical effects.

The rest of this paper is organized as follows: Sec.~\ref{sec:model} describes the various waveform models in the \phX family, phasing coefficients in the \phXAS model, parametrized waveform in the TIGER framework, and the setup of a Bayesian analysis framework to perform the parametrized test; Sec.~\ref{sec:systematic} demonstrates how missing physics like spin precession or higher-order modes in the TIGER can lead to false deviation from GR and our new implementation alleviates this issue; Sec.\ref{sec:GWTC-3} presents the analyses of the GWTC-3 events and includes a separate discussion of two notable asymmetric binary mergers, GW190412 and GW190814. Finally, we conclude in Sec.~\ref{sec:conclusion}.

\section{Waveform model and parametrized test of general relativity}
\label{sec:model}

We develop the improved TIGER framework based on the quasi-circular \phX waveform family. This new framework includes two classes of waveform models: one for binary black holes (\phXAS \cite{Pratten:2020fqn}, \phXP~\cite{Pratten:2020ceb}, \phXHM~\cite{Garcia-Quiros:2020qpx}, \phXPHM~\cite{Pratten:2020ceb}) and another for binary neutron stars (\phXAST~\cite{Colleoni:2023czp}, \phXPT~\cite{Colleoni:2023czp}, \phXASTv~\cite{Abac:2023ujg}, and \phXPTv~\cite{Abac:2023ujg}). The \phXAS model describes the dominant $\ell = 2, \: \abs{m} = 2$ spherical harmonic mode of non-precessing binary. The extension of that model is \phXHM, which contains all the potential higher harmonics $(\ell, \abs{m})=(2,1),(3,3),(3,2),(4,4)$, and mode mixing effects for the (3,2) spherical harmonic.  The \phXPHM model is an extension of \phXHM to describe the precessing binaries based on the technique of ``twisting up'' the non-precessing waveform. This technique maps the aligned-spin waveform modes in the co-precessing frame to the precessing waveform modes in the inertial frame. The \phXP model is a particular case of \phXPHM in that it allows the dominant quadrupole contribution in the co-precessing frame. For binary neutron star models, the tidal corrections are added to the aligned-spin model \phXAS, which is known as \phXAST model. Its precession version \phXPT describes the matter effects in the twisting-up framework. The most recent model \NRTidalvThree improves upon previous semi-analytical BNS model \NRTidalvTwo by calibrating with a larger set of NR data and including dynamical tidal effects, thereby providing more accurate modelling for high-mass-ratio systems and covering a wider range of equations of state.

Since \phXAS serves as the baseline model for all the models in the \phX waveform family, the parametrized deviation parameters are integrated into the phase coefficients of this baseline model. We follow a similar treatment as introduced in the previous studies with \phP waveform family~\cite{Meidam:2017dgf, LIGOScientific:2016lio}.

%IMRPhenomXP describes the precessing binary containing only quadrupolar modes. The IMRPhenomXHM and IMRPhenomXPHM both the models 

\subsection{\phX phasing model}
The frequency domain phasing of the \phXAS model comprises three regimes~\cite{Pratten:2020fqn}: inspiral, intermediate, and merger-ringdown. In the phenomenological model, these regimes are implemented separately using different physical approaches: a PN formalism for the low-frequency inspiral, a quasi-normal ringdown approach for high frequencies, and an intermediate regime that bridges the low and high-frequency evolution by calibrating with NR simulations. Finally, these three segments are smoothly connected by ensuring the $\mathcal{C}^1$ continuous condition in both phase and amplitude. For explicit details of the implementation, we refer the reader to~\cite{Pratten:2020fqn, lalsuite}; here, we only describe the phase, which is primarily used to design the parameterized test.
\begin{enumerate}
\item \emph{Inspiral regime}: A low-frequency inspiral regime is described through PN expansion, with additional terms included at higher PN orders. The PN exapansion is parameterized by PN coefficients $\left\{\varphi_0,...,\varphi_7 \right\}$ and $\left\{\varphi_{5}^{(\ell)}, \varphi_{6}^{(\ell)} \right\}$. The higher-order additional terms are pseudo-PN coefficients $\left\{\sigma_0,...,\sigma_5 \right\}$ that appear between the 4PN and 6PN orders. The full inspiral phase is written as 
\begin{equation}
\label{eq:InsPhase}
\begin{split}
& \Phi_{\rm Ins} = 2\pi f t_c - \varphi_c - \frac{\pi}{4} + \\
 & \underbrace{\frac{3}{128\eta} \sum_{i=0}^7 \left[ \varphi_i(\vec{\lambda}) + \varphi_{i}^{(\ell)}(\vec{\lambda})\log{\left(\pi M f\right)}\right]
  \left(\pi M f\right)^{(i-5)/3} \ \ }_{\text{\normalsize Standard PN terms}} \\
+ & 
 \frac{1}{\eta} \Bigg( \sigma_0 + \sigma_1 f \\ & \underbrace{ \ \ + \frac{3}{4} \sigma_2 f^{4/3}   + \frac{3}{5} \sigma_3 f^{5/3}  + \frac{1}{2} \sigma_4 f^{6/3} + \frac{3}{7} \sigma_5 f^{7/3} \Bigg)  }_{\text{\normalsize Pseudo-PN terms}},
\end{split}
\end{equation}
where $t_c$ and $\phi_c$ reference time and reference phase, respectively. The parameter vector $\vec{\lambda}$ denotes the source parameters component masses ($m_1, m_2$) and spins ($\chi_1, \chi_2$).  The inspiral phase ends at the minimum energy circular orbit (MECO) frequency $\fmeco$, the orbit at which the orbital energy is at its lowest value~\cite{Cabero:2016ayq}.

\item \emph{Intermediate regime:} An intermediate regime describes the dynamics of the binary during the merger phase and serves as a bridge between the inspiral and ringdown phases. In the phenomenological model, this regime is characterized by a polynomial ansatz along with an additional Lorentzian term. The polynomial term is parameterized by a set of phenomenological coefficients $\left\{ b_0,...,b_4 \right\}$. The intermediate phase is written as 
\begin{align}
\label{eq:IntPhase}
 \Phi_{\rm Int} &=\frac{1}{\eta}\Bigg\{b'_0 + b_{0}f + b_1 \log{f} - b_2 f^{-1} -  \frac{b_3}{2} f^{-2} \\
 &\qquad \qquad  -\frac{b_4}{3} f^{-3} \nonumber - \frac{2 c_L}{\fdamp} \tan^{-1}\left(\frac{f- \frd}{2\fdamp} \right) \Bigg\},
\end{align}
where $\frd$ denotes the ringdown frequency, which approximately corresponds to the peak of the Lorentzian. The quantity $\fdamp$ denotes the damping frequency, which corresponds to the width of the Lorentzian.
\item \emph{Merger-Ringdown regime:} A high-frequency regime where the waveform is dominated by quasi-normal ringdown,
\begin{align}
\label{eq:MRPhase}
\Phi_{\rm MR} & = \frac{1}{\eta}\Bigg\{c'_0 + c_0 f + \frac{3c_1}{2} f^{2/3} - c_2 f^{-1} \nonumber \\
& \qquad - \frac{c_4}{3} f^{-3} + \frac{c_L}{\fdamp}  \tan^{-1}\left(\frac{f- \frd}{\fdamp} \right) \Bigg\},
\end{align}
where the negative powers of frequency are added to account for the steep gradient in the inspiral phase.

\end{enumerate}

For comparable-mass binaries, the standard termination frequency for the inspiral phase is at \( \fmeco \). However, when generating waveforms, the transition frequency between the inspiral and intermediate regimes is slightly adjusted for numerical purposes. The termination frequency for the inspiral phase is given by:
\begin{equation}
f_{\rm{Ins}}^{\Phi} = \fmeco - \delta_{R},
\end{equation}
where \( \delta_{R} = 0.03 \left( 0.3 \frd + 0.6 \fisco - \fmeco \right) \). Here, \( \fisco \) represents the frequency at the innermost stable circular orbit (ISCO).

The transition frequency between the intermediate and merger-ringdown phases is given by:
\begin{equation}
f_{\rm{Int}}^{\Phi} = 0.6 \left( 0.5 \frd + \fisco \right) + \frac{1}{2}\delta_{R}.
\end{equation}
It is important to note that the transition frequencies for the amplitude differ from those for the phase~\cite{Pratten:2020fqn}.

%\subsection{Effect of testing parameter}
\subsection{Parametrized waveform in the TIGER approach}
Within the TIGER framework, we construct the parametrized waveform model by incorporating a fractional deviation parameter ($\dpi_i$) into the phasing coefficients,
\begin{equation}
\phi_i^{\mathrm{GR}} \rightarrow (1+\dpi_i)\: \phi_i^{\mathrm{GR}},
\end{equation}
where $\phi_i^{\mathrm{GR}}$ represents the PhenomX phasing coefficients 
\begin{equation}
\label{eq:phiGR}
\phi_i^{\mathrm{GR}} \equiv \big\{ \underbrace{ \varphi_0,...,\varphi_7,  \varphi_{5}^{(\ell)}, \varphi_{6}^{(\ell)}}_{\text{Inspiral}}, \  \underbrace{ b_1, b_2, b_3, b_4}_{\text{Intermediate}}, \ \underbrace{ c_1, c_2, c_4, c_L}_{\text{Merger-ringdown}}\big\}.
\end{equation}
We note that the deviations in the inspiral regime are introduced only in the non-spinning part of the PN coefficients, specifically for $\varphi_3$, $\varphi_4$, $\varphi_5^{(\ell)}$, $\varphi_6$, and $\varphi_7$. However, our TIGER implementation in \lalsuite software\footnote{There was an issue in the TIGER implementation in the \lalsuite software for the –1PN and –0.5PN cases. Please use this checkout \href{https://git.ligo.org/lscsoft/lalsuite/-/commits/653065c0f77c76f413aae3b09c41b3b878c11840}{653065c0f77c76f413aae3b09c41b3b878c11840} or the \texttt{master} branch until a new tag is released.} supports deformations in either the spinning or non-spinning part of the PN coefficients, or both equally~\cite{lalsuite}.
 
We exclude the terms with unit power of frequency and the non-logarithmic term at 2.5PN since they can be absorbed in a redefinition of $t_c$ and $\varphi_c$, respectively. We also exclude the pseudo-PN terms that enter beyond 3.5 PN and are used for calibrating with the numerical waveform data set. The pairs $(b'_0, b_0)$ and $(c'_0, c_0)$ are also excluded as they are used to ensure the $\mathcal{C}^1$ continuity condition of the waveform.

Many alternative theories of gravity include scalar fields alongside tensor fields, resulting in a leading dipolar contribution at -1PN and a tail-induced dipole contribution at 0.5PN~\cite{Will:1994fb, Yunes:2011aa, Arun:2012hf, Barausse:2016eii, Yagi:2015oca}. In contrast, solely GR does not permit these parameters. To search for the non-GR dipolar radiation, we introduce the absolute deviation parameters at -1PN and 0.5PN. The correction to the GW phase is given by,
\begin{equation}
\Delta\Phi = \dphi_i \frac{3}{128\eta} (\pi M f)^{(i-5)/3},
\end{equation}
where  $i = -2$  and  $i = 1$  correspond to the -1PN and 0.5PN terms, respectively.

%The inspiral phase of higher-order modes in the time domain is modeled by scaling the dominant quadrupole mode, which allows any deviations in the inspiral phase to propagate to the higher-order modes.

\subsubsection{Parametrized deviation into higher-order modes}
%The higher-order modes of gravitational wave signals from coalescing compact binaries play a vital role in the accurate reconstruction of source properties. Consequently, it allows for stronger constraints on the deviation from GR since the deviation parameters are highly correlated with chirp mass. In contrast, the absence of higher-order modes in the parametrized framework can lead to false deviation of GR if their contribution is large in the signal. 

The GW wave signal $h(t)$, propagating along an arbitrary direction $(\iota,\phi_0)$ in the source frame, can be decomposed over the spin-weighted spherical harmonic basis (with spin-weight -2) as:
\begin{equation}
h(t; \vec{\lambda}, \iota, \phi_0) = \sum_{\ell=2}^{\infty}\sum_{m=-\ell}^{m=\ell} h_{\ell m}(t; \vec{\lambda}) \; {}_{-2}Y_{\ell m}(\iota, \phi_0),
\end{equation}
where $h_{\ell m}(t; \vec{\lambda}) = A_{\ell m}(t; \vec{\lambda}) \; e^{i\Phi_{\ell m}(t; \vec{\lambda})}$ represents the $(\ell, m)$ mode described by the amplitude $A_{\ell m}(t; \vec{\lambda})$ and phase $\Phi_{\ell m}(t; \vec{\lambda})$. In particular, for non-precessing spinning BHs, the inspiral phase of an arbitrary $(\ell, m)$ mode can be expressed in terms of the phase of the $(2,2)$ mode alone: ${ \Phi_{\ell m}(t) \simeq   (m/2) \; \Phi_{22}(t) }$. At a given PN order, if a phase correction is required for the $(2, 2)$ mode to accurately represent the signal, the same correction must be included in the $(\ell, m)$ mode phase, scaled by a factor of $m/2$. This suggests that the deviation introduced to the dominant mode should also be propagated to all higher-order modes. As we construct our parametrized model in the frequency domain, it is convenient to express the multimode as:
\begin{equation}
\tilde{h}_{\ell m}(f) = A_{\ell m}(f)\; e^{i\Phi_{\ell m}(f)}.
\end{equation}
For the inspiral phase, the relation among the modes in the time domain can be expressed in the Fourier domain as,
\begin{equation}
\Phi^{\mathrm{Ins}}_{\ell m}(f) \approx \frac{m}{2} \Phi^{\mathrm{Ins}}_{22}\left( \frac{2}{m} f\right),
\end{equation}
where $\Phi^{\mathrm{Ins}}_{22}$ is given in Eq.~\eqref{eq:InsPhase}. This relation allows us to propagate the same deviation in PN coefficients into the higher-order modes. The non-GR correction to the inspiral phase for $(\ell, m)$ mode is expressed as,
\begin{align}
\label{eq:dphilm}
\Delta\Phi_{\ell m} &= \frac{3}{128\eta} \left(\frac{2\pi M f}{m}\right)^{(k-5)/3} \times \nonumber \\
&\begin{cases}
 \delta\hat{\varphi}_{k}^{(\ell)} \: \varphi_{k}^{(\ell)}(\vec{\lambda}) \: \log{\left(\frac{2\pi M f}{m}\right)}, & \text{for Log terms}, \\
 \delta\hat{\varphi}_k , & \text{for $k=-2, 1$}, \\
\delta\hat{\varphi}_k \: \varphi_k(\vec{\lambda}), & \text{otherwise}
\end{cases}
\end{align}
with the superscript $(\ell)$ denote the coefficients of the log terms in PN expansion, not related to the $(\ell, m)$ mode.

The scaling relation between the modes is only valid during the inspiral phase and does not extend to the intermediate and merger-ringdown stages. In the \phXHM model, the phenomenological coefficients for the post-inspiral part are modeled separately for individual modes. As a result, deviations in the post-inspiral phase do not affect the higher-order modes. No post-inspiral deviation parameters are introduced for the higher-order modes in this new TIGER framework.

%The multimodes of a frequency domain waveform from GW signal from non-precessing binary is expressed as,
%\begin{equation}
%\tilde{h}_{\ell, m}
%\end{equation}

\subsubsection{Twisting up the parametrized deviation}
In the context of modeling GWs from precessing spin binaries, ``twisting up'' is a technique to map between non-precessing waveform modes in the co-precessing frame and precessing waveform modes in the inertial frame~\cite{Schmidt:2012rh}. When the spin components of the BHs are misaligned with the orbital angular momentum, the binary plane experiences a general relativistic precession and nutation, which results in an observable modulation in the phase and amplitude of the GW signal~\cite{Apostolatos:1994mx, Kidder:1995zr}. The twisting up technique approximates amplitude and phase modulation by performing a time-dependent rotation over the non-precessing modes~\cite{Schmidt:2010it}.

To outline the procedure, we consider the $L$-frame (also known as the co-precessing frame), where the $z$-axis is aligned with the orbital angular momentum, and the $J$-frame (the inertial frame), where the $z$-axis is aligned with the total angular momentum. The Euler angles $(\alpha, \beta, \gamma)$ are defined to represent an active rotation from the inertial $J$-frame to the precessing $L$-frame. The GW modes in these two frames are related through the transformation of a Weyl scalar under a rotation $\mathcal{R} \in \mathrm{SO}(3)$,
\begin{align}
    h^{J}_{\ell m} & = \sum_{m'=-\ell}^{\ell} \mathcal{D}^{\ell*}_{mm'}(\alpha,\beta,\gamma) h^{L}_{\ell m'}, \\
    h^{L}_{\ell m'} & = \sum_{m=-\ell}^{\ell} \mathcal{D}^{\ell}_{mm'}(\alpha,\beta,\gamma) h^{J}_{\ell m},
\end{align}
where $\mathcal{D}^{\ell*}_{mm'}(\alpha,\beta,\gamma)$ are the Wigner D-matrices,
\begin{equation}
    \mathcal{D}^{\ell}_{mm'}(\alpha, \beta, \gamma) = e^{im\alpha} e^{im'\gamma} d^{\ell}_{mm'}(\beta),
\end{equation}
where $d^{\ell}_{mm'}(\beta)$ are the real-valued Wigner-d matrices. The frequency-domain expressions for the GW polarizations in the inertial $J$-frame $\tilde{h}_{+,\times}^J (f)$ in terms of spherical harmonic modes $\tilde{h}_{\ell m}^L (f)$ in the co-precessing $L$-frame are given as,
%\begin{align}
%\label{eq:polarizations_1}
%    \tilde{h}_{+}^{J} (f>0) &= \frac{1}{2} \displaystyle\sum_{\ell \geq 2} \displaystyle\sum_{m^{\prime} > 0}^{l} \tilde{h}^{L}_{\ell  -m^{\prime}} (f) e^{ i \mpr \gamma} \displaystyle\sum_{m=-\ell}^{\ell}\left[ A^{\ell}_{m \, -\mpr} + (-1)^{\ell} A^{\ell \, \ast}_{m \, \mpr} \right], 
%    \\
%    \label{eq:polarizations_2}
%    \tilde{h}_{\times}^{J} (f>0) &= \frac{i}{2} \displaystyle\sum_{\ell \geq 2} \displaystyle\sum_{\mpr > 0}^{\ell} \tilde{h}^{L}_{\ell -\mpr } (f) e^{i \: \mpr \gamma} \displaystyle\sum_{m=-\ell}^{\ell} \left[ A^{\ell}_{m - \mpr} - (-1)^{\ell} A^{\ell\: \ast}_{m \mpr} \right],
%\end{align}
\begin{align}
\label{eq:polarizations}
    &\tilde{h}_{\ptimes}^{J} (f>0) = \frac{1}{2} \displaystyle\sum_{\ell \geq 2} \displaystyle\sum_{m^{\prime} > 0}^{\ell} \tilde{h}^{L}_{\ell  -m^{\prime}} (f) e^{ i \mpr \gamma} \\ \nonumber
    & \times \sum_{m=-\ell}^{\ell}\left[ e^{-i m \alpha} d^{\ell}_{m -\mpr}(\beta) \; _{-2}Y_{\ell m} \: \pm \: (-1)^{\ell} e^{-i m \alpha} d^{\ell}_{m \mpr}(\beta) \; _{-2}Y_{\ell m}^{\ast} \right],
\end{align}
%where $ A^{\ell}_{m \, \mpr}$ is related to Wigner-$d$ matrices, 
%\begin{equation}
%   A^{\ell}_{m \, \mpr} =  e^{-i \: m \:\alpha} d^{\ell}_{m \mpr}(\beta) \; _{-2}Y_{\ell m}. 
%\end{equation}
where the plus-minus symbol $\pm$ corresponds to $\ptimes$ polarizations, respectively. The modes $\tilde{h}_{\ell m}$ and $\tilde{h}_{\ell -m}$ are connected by equatorial symmetry, expressed as $\tilde{h}_{\ell m}(f) = (-1)^{\ell} \tilde{h}^{\ast}_{\ell -m}(-f)$, which is only valid for non-precessing binaries. Here, the modes are $(\ell, \abs{m})=(2,1),(3,3),(3,2),(4,4)$, as implemented in the \phXHM model~\cite{Garcia-Quiros:2020qpx}. For complete details on this twisting-up technique for \phXPHM model, we refer the reader to~\cite{Pratten:2020ceb}.

In the TIGER framework, we apply the phase correction in the $L$-frame, which does not affect the spin dynamics of the binary. This implies that Euler angles are completely determined by the GR parameters and are not changed by the parametrized deviation. However, our beyond-GR correction in $L$-frame,
\begin{equation}
\Delta\tilde{h}_{\ell m} = \tilde{h}_{\ell m} \left( e^{i\Delta\Phi_{\ell m}} -1 \right)
\end{equation}
is twisted up to obtain the waveforms in $J$-frame, where $\Delta\Phi_{\ell m}$ is given in Eq.~\eqref{eq:dphilm}. Therefore, the beyond-GR correction in the TIGER framework can be obtained by substituting $\Delta\tilde{h}_{\ell m}$ into Eq.~\eqref{eq:polarizations}, yielding:
\begin{align}
\label{eq:polarizations_d}
    &\Delta\tilde{h}_{\ptimes}^{J} (f>0) = \frac{1}{2} \displaystyle\sum_{\ell \geq 2} \displaystyle\sum_{m^{\prime} > 0}^{\ell} \Delta\tilde{h}^{L}_{\ell  -m^{\prime}} (f) e^{ i \mpr \gamma} \\ \nonumber
    & \times \sum_{m=-\ell}^{\ell}\left[ e^{-i m \alpha} d^{\ell}_{m -\mpr}(\beta) \; _{-2}Y_{\ell m} \: \pm \: (-1)^{\ell} e^{-i m \alpha} d^{\ell}_{m \mpr}(\beta) \; _{-2}Y_{\ell m}^{\ast} \right]
\end{align}
This expression shows how the parametrized deviations in the TIGER framework are twisted-up when accounting for the frame transformation effects.

\subsubsection{Connection with \phP model}
In the previous version of TIGER with the \phP family~\cite{Hannam:2013oca, LIGO_T1500602, Khan:2019kot}, parametrized deviation terms were incorporated into the phasing coefficients of the \phD model~\cite{Husa:2015iqa, Khan:2015jqa}. The inspiral phasing for both the \phD and \phXAS models is based on the TaylorF2 approximation, derived using the stationary phase approximation, and its functional form is provided in Eq.~\eqref{eq:InsPhase}.
For the post-inspiral phase, the \phX model employs a similar phasing approach as the \phD model. However, there are differences in the phasing coefficients, their frequency powers, and the transition frequencies between the three frequency regimes.
%In \phD model, the transition frequencies are determined based on the total mass of the binary ($M$) and the remnant mass and spin. The transition frequency from inspiral to intermediate regime is given by the condition $GMf/c^3 = 0.018$. The transition between intermediate and merger-ringdown in determined by the condition $GMf/c^3 = 0.5 f_{\mathrm{RD}}$

The intermediate phasing of the \phD model is given in Eq.\:(16) of Ref.~\cite{Khan:2015jqa} as:
\begin{equation}
\Phi_{\mathrm{Int}} = \frac{1}{\eta} \left\{ \beta_0 + \beta_1 f + \beta_2 \log(f) - \frac{\beta_3}{3}f^{-3} \right\},
\end{equation}
where $\beta_0,...,\beta_3$ are phenomenological coefficients. In the previous TIGER framework, the parametrized deviations were introduced into $\beta_2$ and $\beta_3$. In the \phXAS phasing, as shown in Eq.~\eqref{eq:IntPhase}, there are two analogous terms with the same frequency power, where $\beta_2 \leftrightarrow b_1$ and $\beta_3 \leftrightarrow b_4$.

The merger-ringdown phase of \phD model is given in Eq.\:(14) of Ref.~\cite{Khan:2015jqa} as:
\begin{align}
\Phi_{\rm MR} & = \frac{1}{\eta}\Bigg\{\alpha_0 + \alpha_1 f - \alpha_2 f^{-1} \nonumber \\
& \qquad + \frac{4}{3} \alpha_3 f^{3/4} + \alpha_4  \tan^{-1}\left(\frac{f- \alpha_5\frd}{\fdamp} \right) \Bigg\},
\end{align}
where $\alpha_0,...,\alpha_5$ are phenomenological coefficients. In the previous TIGER framework, the parametrized deviations were introduced into $\alpha_2$, $\alpha_3$, and $\alpha_4$. In \phXAS phasing, shown in Eq.~\eqref{eq:MRPhase}, corresponding terms appear at the same frequency power, with $\alpha_2 \leftrightarrow c_2$.

In both models, there is a subset of coefficients that enter at the same frequency power, but the modeling differs. The \phD coefficients are obtained by fitting a 2D parameter space of symmetric mass ratio $\eta\equiv m_1 m_2/(m_1+m_2)^2$ and effective spin $\chi_{\rm eff}\equiv m_1\chi_1 +  m_2\chi_2/(m_1+m_2)^2$, with different effective spins for different frequency regimes as needed. In contrast, \phXAS coefficients are derived by fitting a 3D parameter space, still expressed in terms of the symmetric mass ratio, effective spin, and spin difference $\Delta\chi = \chi_1 - \chi_2$. The additional term $\Delta\chi$ accounts for the unequal spin contribution. In Sec.~\ref{sec:GWTC-3}, we did not include the \phP-based results of GWTC-1/2 events for the post-inspiral parameters in our combined results with GWTC-3 events, though we did include them for the inspiral parameters.

\subsection{Bayesian analysis framework}
The parametrized test with \phX family described above introduces an additional set of non-GR parameters,
\begin{align}
\dpi_i \equiv & \Big\{\overbrace{ \delta\hat{\varphi}_{-2}, \delta\hat{\varphi}_0,..., \delta\hat{\varphi}_7,    \delta\hat{\varphi}_{5}^{(\ell)}, \delta\hat{\varphi}_{6}^{(\ell)}}^{\text{\normalsize Inspiral}},    \quad \overbrace{ \delta\hat{b}_1, \delta\hat{b}_2, \delta\hat{b}_3, \delta\hat{b}_4}^{\text{\normalsize Intermediate}}, \nonumber \\
& \quad \quad \quad \quad \quad \quad  \underbrace{\delta\hat{c}_1, \delta\hat{c}_2, \delta\hat{c}_4, \delta\hat{c}_L}_{\text{\normalsize Merger-ringdown}}\Big\},
\end{align}
which correspond to the phase coefficients of the \phXAS model as given in Eq.~\eqref{eq:phiGR}. Our parametrized test involves a nested hypothesis where the GR is a special case. The beyond-GR model is characterized by all the standard GR signal parameters ($\vec{\theta}_{\mathrm{GR}}$) plus one additional non-GR parameter \mbox{$\dpi_i$, such that $\vec{\theta} \equiv \{\vec{\theta}_{\mathrm{GR}}, \dpi_i \}$}. For a given data $d$ and hypothesis $\mathcal{H}$, we obtain the posterior probability density \mbox{$p\big(\vec{\theta} \mid d,\mathcal{H}\big)$} using Bayes' theorem,
\begin{equation}
\label{eq:posterior}
p\big(\vec{\theta} \mid d, \mathcal{H} \big) \propto p\big(d \mid \vec{\theta}, \mathcal{H} \big) \; p\big(\vec{\theta} \mid \mathcal{H} \big),
\end{equation}
where \mbox{$p\big(\vec{\theta} \mid \mathcal{H} \big)$} is the prior probability distribution, and \mbox{$p\big(d \mid \vec{\theta}, \mathcal{H} \big)$} is the likelihood. We use the standard likelihood function assuming additive noise that is stationary and Gaussian~\cite{Veitch:2009hd},
\begin{equation}
p\big(d \mid \vec{\theta}, \mathcal{H} \big) \propto \exp\left[ \frac{1}{2} \left< d - h(\vec{\theta}) \mid d - h(\vec{\theta}) \right>  \right],
\end{equation}
where the angular brackets, $\inp{\cdot}{\cdot}$, represent the noise-weighted inner product, defined as
\begin{equation}
 \inp{a}{b} \equiv 4 \Re \int_{\flow}^{\fhigh}  df \, \frac{ \tilde{a}^{\ast}(f) \, \tilde{b}(f) }{S_n (f)} \,, 
\end{equation}
where $S_n(f)$ is the one-sided power spectral density (PSD) of the detector noise, and $ \tilde{a}(f) $ and $\tilde{b}(f)$ denote the Fourier transform of two time-series signals $a(t)$ and $b(t)$, respectively. The integration limits \flow and \fhigh represent the lowest and highest cutoff frequencies of the sensitivity bandwidth. We analyzed all LVK events with a lower cutoff frequency of 20\Hz. This specific configuration was selected based on information available in the LVK public data. We estimate the beyond-GR effects by computing the marginalized posterior probability distribution of the $\dpi_i$, integrating the posterior distribution \mbox{$p(\vec{\theta} \mid d,\mathcal{H})$} over the nuisance parameters,
\begin{equation}
\label{eq:dpi_posterior}
    p\left(\dpi_i \,|\, d \right) = \int \left( \prod_{\vec{\theta} \,\setminus \{\dpi_i\}} d\theta_i \right) \: p( \vec{\theta} \,|\, d,  \mathcal{H} ).
\end{equation}

To obtain the posterior probability distribution \mbox{$p(\vec{\theta} \mid d,\mathcal{H})$} in Eq.~\eqref{eq:posterior}, we typically use the algorithms such as Markov-chain Monte Carlo or Nested Sampling. Throughout this work, we use the \bilbytgr package~\cite{bilby-tgr} based on \bilby~\citep{Ashton:2018jfp}, with the nested sampling algorithm \dynesty sampler~\cite{Speagle:2019ivv}.

For the prior probability distribution \mbox{$p\big(\vec{\theta} \mid \mathcal{H} \big)$}, we follow the setup provided in the LVK public data. For the GR parameters, we assume a uniform prior on component masses, uniform prior on individual spin magnitudes with isotropic orientations, and a luminosity distance $d_L$ prior that is uniform in comoving volume, given by $p(d_L) \propto d_L^2$. For the non-GR parameter, we assume a uniform prior symmetric around zero.

To assess the effectiveness of the different PN deviation coefficients at spotting deviation from GR or level of consistency with GR, we compute the fraction of the posterior enclosed by the smallest  Highest Posterior Density (HPD) interval that contains the GR value. We refer to this quantity as the GR quantile, denoted as $\QGR$. A GR quantile value of 0\% indicates that the GR value lies exactly at the peak of the posterior (perfect consistency with GR), whereas a value of 100\% means that the GR value lies at the edge or outside the 100\% credible region, implying a strong deviation from GR. An intermediate value of $\QGR$ quantifies the level of consistency between the GR prediction and the recovered posterior of the deviation parameter.
%the smallest Highest Posterior Density (HPD) interval that encompasses the GR value. We denote this value as $\QGR$. %For a 1D posterior, it is expressed as,
%\begin{equation}
%\label{eq:qgr}
%\QGR(\dpi_i) = \:\mid 2 P(\dpi_i<0) -1 \mid
%\end{equation}
%where $P(\dpi_i<0)$ represents the fraction of posterior samples below 0. %For 2D posterior distribution, we find the isoprobability contour that encompasses GR. Let assume $p(x, y)$ represents the probability density function for the two variables $x$ and $y$. We find the isoprobability contour $R$, where $(0, 0)\in R$ and then $\QGR$ is expressed as,
%\begin{equation}
%\QGR = \iint_R p(x, y) \, dx \, dy
%\end{equation}

To compare two competing hypotheses, we compute the ratio of their evidences, referred to as the Bayes factor. Assuming we compare the GR hypothesis ($\mathcal{H}_{\rm GR}$) against one incorporating non-GR effects ($\mathcal{H}_{\rm nonGR}$), the Bayes factor is defined as
\begin{equation}
\mathcal{B}_{\rm GR}^{\rm nonGR} = \frac{p\big(d \mid \mathcal{H}_{\rm nonGR}\big)}{p\big(d \mid \mathcal{H}_{\rm GR}\big)},
\end{equation}
where $p\big(d \mid \mathcal{H}_{\rm nonGR}\big)$ and $p\big(d \mid \mathcal{H}_{\rm GR}\big)$ are the evidences for $\mathcal{H}_{\rm nonGR}$ and $\mathcal{H}_{\rm GR}$, respectively. We typically compute the evidence by integrating the likelihood $p\big(d \mid \vec{\theta}, \mathcal{H} \big)$ weighted by the prior $p\left(\vec{\theta}\mid \mathcal{H}\right)$ over the entire parameter space:
\begin{equation}
p\big(d \mid \mathcal{H}\big) = \int p\big(d \mid \vec{\theta}, \mathcal{H} \big) \: p\big(\vec{\theta}\mid \mathcal{H}\big) \: d\vec{\theta}.
\end{equation}
To ensure that the posterior distribution in Eq.~\eqref{eq:posterior} is a valid probability distribution, we normalize the posterior by evidence.

\section{Impact of higher harmonics and precession}
\label{sec:systematic}
In this section, we demonstrate the impact of higher harmonics and precession when one of these components is omitted from the baseline GR waveform model in TIGER. We consider a set of injections that includes significant contributions from higher-order modes and binary precession effects. The parameterized tests are then performed using the \phXHM and \phXP models. The \phXHM model includes higher-order modes but excludes precession, while the \phXP model incorporates precession but omits higher-order modes.
\begin{table}
    \centering
    \setlength{\tabcolsep}{1pt}
    \begin{tabularx}{0.49\textwidth}{l c c c}
        \toprule[1pt]
        \toprule[1pt]
        Parameters  & Case I & Case II & Case III \\
        \midrule[0.5pt]
        Total mass $[\msun]$  & 78.5  & 78.5 & 78.5  \\
        Mass ratio ($m_2/m_1$) & 0.42 & 0.42 & 1/9 \\
        Primary spin ($a_1, \theta_1$) & (0.99, 1.42) & (0.99, 1.42) & (0, 0) \\
        Secondary spin ($a_2, \theta_2$) & (0.6, 2.0) & (0.6, 2.0) & (0, 0) \\
        Azimuthal angles ($\phi_{12}$, $\phi_{\mathrm{JL}}$) & (5.3, 1.4) & (5.3, 1.4) & -- \\
        Zenith angle ($\theta_{\mathrm{JN}}$) & 0.4 & $\pi/2$ & $\pi/2$ \\
        Luminosity distance ($d_L$ [Mpc]) & 910 & 910 & 364.0 \\
        
        \midrule
        
        Waveform model & \multicolumn{3}{c}{Log Bayes factor values}\\
        &  \multicolumn{3}{c}{(\logB{}) for GR template} \\
        \midrule
        \phXPHM (\logB{XPHM}) & 417.9 & 101.1 & 119.2 \\
        \phXP (\logB{XP}) & 413.6 & 94.9 & 83.0 \\
        \phXHM (\logB{XHM}) & 405.1 & 57.1 & 119.6  \\
        \bottomrule[1pt]
        \bottomrule[1pt]
    \end{tabularx}
    \caption{Parameter values for GW200129-like injections, where Case I corresponds to the maximum likelihood point from event parameter estimation with \phXPHM model. Details on the spin parametrization are available in the \lalsimulation documentation~\cite{LALSimulation_inference}. The other extrinsic parameters are fixed for all the cases as follows: polarization angle of 2.3, geocentric GPS time of 1264316116.4 seconds, sky location at right ascension 5.6 and declination 0.04, and a coalescence phase of 0.41. The bottom panel shows the logarithmic Bayes factor values for three different models, where the injections were generated using the \phXPHM model.  
    }
    \label{tab:GW200129maxL}
\end{table}

%\begin{table}[t]
%    \centering
%    \setlength{\tabcolsep}{3pt}
%    \begin{tabularx}{0.46\textwidth}{X c c c}
%        \toprule[1pt]
%        \toprule[1pt]
%        Parameters  & Case-I & Case-II  & Case-III \\
%        \midrule[0.5pt]
%        Total mass $[\msun]$  & 78.5  & 78.5 & 78.5  \\
%        Mass ratio ($m_2/m_2$) & 0.42 & 0.42 & 1/9 \\
%        Primary spin ($a_1, \theta_1$) & (0.99, 1.42) & (0.99, 1.42) & (0, 0) \\
%        Secondary spin ($a_2, \theta_2$) & (0.6, 2.0) & (0.6, 2.0) & (0, 0) \\
%        Azimuthal angles ($\phi_{12}$, $\phi_{\mathrm{JL}}$) & (5.3, 1.4) & (5.3, 1.4) & (0, 0) \\
%        Zenith angle ($\theta_{\mathrm{JN}}$) & 0.4 & $\pi/2$ & $\pi/2$ \\
%        \bottomrule[1pt]
%        \bottomrule[1pt]
%    \end{tabularx}
%    \caption{Parameter values for GW200129-like injections, where Case-I corresponds to the maximum likelihood point from event parameter estimation with \phXPHM model. Details on the spin parametrization are available in the \lalsimulation documentation.~\cite{LALSimulation_inference}. The other extrinsic parameters are fixed for all the cases as follows: a luminosity distance of 910 Mpc, polarization angle of 2.3, geocentric time of 1264316116.4 GPS, sky location at right ascension 5.6 and declination 0.04, and a coalescence phase of 0.41.}
%    \label{tab:GW200129maxL}
%\end{table}
\begin{figure*}
    \centering
    \includegraphics[width=0.98\textwidth]{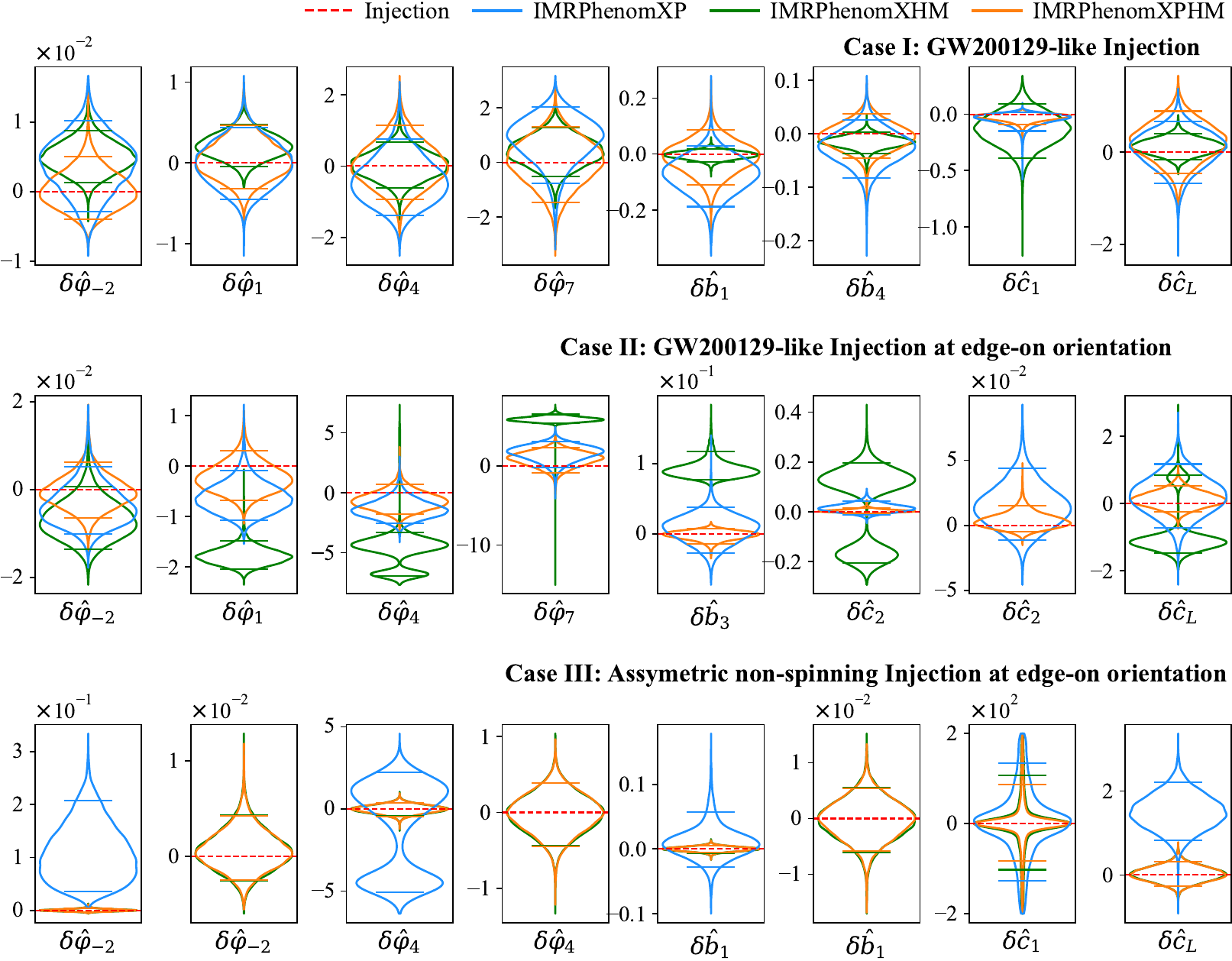}
    \caption{Systematic errors and false deviation in the parametrized test of GR with TIGER due to missing physics, shown for GW200129-like injections; corresponding parameter values are listed in Table~\ref{tab:GW200129maxL}. The GR injection (shown in red) was generated using a precessing spin with higher-order modes model \phXPHM, and recovered with precessing spin without higher-order modes model \phXP (blue), non-precessing spin with higher-order modes model \phXHM (green), and \phXPHM (orange) model, which includes both. The small horizontal lines represent the 90\% credible interval of the posteriors. In the middle and bottom rows, for some deviation parameters, the posteriors obtained using \phXP are significantly broader than those from the other two models. These cases are shown in two consecutive subplots to highlight the difference and to clearly display the narrower posteriors. The full results are shown in Fig.\:\ref{fig:gw200129-like_inj_full}.}
    \label{fig:gw200129-like_inj}
\end{figure*}

To carry out this study, we select a debatable event in \mbox{GWTC-3}, GW200129\_065458 (hereafter GW200129), which was observed as a highly precessing binary merger~\cite{Hannam:2021pit} and recorded the highest SNR among the events in the catalog~\cite{KAGRA:2021vkt}. With this event, Some testing GR pipelines reported false indications of beyond-GR effects~\cite{Maggio:2022hre}. These were attributed either to the use of a non-precessing baseline GR waveform model or data quality issues in the Livingston detector~\cite{Payne:2022spz}.

Here, we only focus on the waveform systematics due to missing physics in the template waveform model. We choose the maximum likelihood point from the posterior samples generated with the \phXPHM model. We consider three cases by varying the parameter values: (i) using the exact maximum likelihood point (Case I), (ii) assuming an edge-on orientation of the binary to achieve the maximum precession effect and a little impact of higher-order modes (Case II), and (iii) a highly asymmetric, non-spinning binary at edge-on orientation to achieve the significant impact of higher-order modes (Case III). The parameter values of these cases are detailed in Table~\ref{tab:GW200129maxL}. To avoid the statistical errors due to noise systematics, we generate the injections assuming a zero-noise realization with a three-detector network of Hanford (H1), Livingston (L1), and Virgo (V1). When performing the parameter estimation runs, we weight the likelihood inner product by the power spectral density corresponding to their advanced designed sensitivities: \texttt{aLIGOZeroDetHighPower} for H1 and L1~\cite{aLIGO_ZDHP}, and \texttt{AdvVirgo} for V1~\cite{2012arXiv1202.4031M}, as implemented in \lalsuite. With this setup, we find injection optimal SNR of 45.1, 23.9, and 25.0 for Case I, II, and III, respectively.

%To access the impact of higher harmonic and precession individually, we report the Bayes factor values (\logB{}) for GR runs with three different models, as listed in the bottom panel of Table~\ref{tab:GW200129maxL}. For Case-I injection, the \logB{XP} is slightly larger than \logB{XHM} and \logB{XPHM} is slightly larger than both, which implies impact of higher harmonics slightly smaller than precession, but not non-negligible. For Case-II injection, the \logB{XP} is significantly larger than \logB{XHM} and \logB{XPHM} is slightly larger than former one, which implies contribution of precession considerably higher than higher harmonic, but the impact of higher harmonics can not ignored. For Case-III injection, the \logB{XHM} and  \logB{XPHM} are nearly equal and significantly larger than \logB{XP}, which implies significant contribution of higher-order modes. Now, we want to look into whether missing of one of them can lead to a biased conclusion.

%In Figure~\ref{fig:gw200129-like_inj}, we present the TIGER results for a GW200129-like injections: the top two rows are shown for the Case-I and bottom two rows are for Case-II, the details of the parameter values are listed in Table~\ref{tab:GW200129maxL}. In all th cases, it is evident that neglecting either higher harmonics or precession in the TIGER waveform model can lead to biased conclusion of deviations from GR.

In Figure~\ref{fig:gw200129-like_inj}, we present the TIGER results for GW200129-like injections listed in Table~\ref{tab:GW200129maxL}. In all the cases, it is evident that neglecting either higher harmonics or precession in the TIGER waveform model can lead to biased conclusions of deviations from GR. In the following paragraph, we report the Bayes factor values (\logB{}) for GR runs with three different models, as listed in the bottom panel of Table~\ref{tab:GW200129maxL}. This will help us assess how the absence of either higher harmonics or precession can lead to a biased conclusion.

\noindent\textbf{Case I:} \logB{XP} is slightly higher than \logB{XHM}, while \logB{XPHM} is marginally greater than both, suggesting that the impact of higher harmonics is slightly less than that of precession, but not inconsiderable. The top row of Fig.~\ref{fig:gw200129-like_inj} presents the results for this injection. In all cases, the injection value lies within the 90\% credible interval, except at -1PN and 0PN for the \phXHM model. The non-inclusion of precession for a GW200129-like event can lead to false deviation from GR. While the impact of higher harmonics is not significant, in some PN terms (2 to 3.5PN), the \phXP posteriors are slightly offset from the injected value.

\noindent\textbf{Case II:} The \logB{XP} is significantly larger than \logB{XHM} and \logB{XPHM} is slightly larger than the former one, which implies the contribution of precession considerably higher than higher harmonics, but the impact of higher harmonics cannot be ignored. The middle row of Fig.~\ref{fig:gw200129-like_inj} presents the results for this injection. The results obtained using \phXHM indicate that the absence of spin precession can lead to a significantly biased conclusion, suggesting false evidence of deviations from GR. For all inspiral parameters ($\delta\hat{\varphi}_i$) except the -1PN, and for all parameters in the intermediate regime ($\delta\hat{b}_i$), the GR prediction lies entirely outside the 99\% credible interval. The exception at -1PN term could be due to its non-correlation with precession, because the leading precession effect enters at 2PN. In the \phXHM results, strong bimodality is seen in the posteriors of the merger-ringdown parameters, with almost no support at the injection value. In the \phXP results, for some lower PN terms (0 to 1PN), the GR prediction falls outside the 90\% credible interval. For higher PN terms, the posteriors are slightly shifted, with the GR prediction close to the 90\% credible interval. For the post-inspiral parameters, the distributions are sufficiently consistent with the GR prediction but significantly broader compared to the \phXPHM model.

\noindent\textbf{Case III:} The values of \logB{XHM} and \logB{XPHM} are nearly identical and significantly higher than \logB{XP}, indicating a substantial contribution from higher-order modes. The bottom row of Fig.~\ref{fig:gw200129-like_inj} shows the results for this injection. In \phXP results, we see strong deviation at -1PN and one merger-ringdown parameter $\delta\hat{c}_\ell$. For all the remaining inspiral parameters, the posteriors are quite broader, have a long tail, and bimodality in higher PN terms starting from 1.5PN. The posteriors of \phXHM and \phXPHM are nearly identical since the injection does not have the precession contribution. This implies that the deviation parameters are not correlated with the precessing spin parameters for non-spinning systems. We note that the merger-ringdown parameters $\delta\hat{c}_1$ and $\delta\hat{c}_4$ show broad posterior support, indicating that they are not well constrained for non-spinning asymmetric binaries. A comparable behavior is seen for the GW190814 event, as shown in Fig.~\ref{fig:violin_gw190412_gw190814}.

%%%%%%%%%%%%%%%%%%%%%%%%%%%%%%%%%%%%%%%%%%%%%%%%%%%%%%%%%%%

\begin{table*}[!t]
\centering
%\vspace{0.3cm}
\setlength{\tabcolsep}{4pt}  % Set column separation
\renewcommand{\arraystretch}{1.15}  % Set row separation 
\begin{tabular}{l c c c c c c @{\hskip 0.2cm}||@{\hskip 0.2cm}l c c c c c c  }
\toprule[1pt]
\toprule[1pt]
Event & $f_c^{\mathrm{PAR}}$ [Hz] & $\rho_{\mathrm{IMR}}$ & $\rho_{\mathrm{insp}}$ & $\rho_{\mathrm{postinsp}}$ & PI & PPI & Event & $f_c^{\mathrm{PAR}}$ [Hz] & $\rho_{\mathrm{IMR}}$ & $\rho_{\mathrm{insp}}$ & $\rho_{\mathrm{postinsp}}$ & PI & PPI \\
\midrule[0.8pt]
\midrule[0.8pt]
\multicolumn{14}{c}{GWTC-1} \\
\hline
GW150914 & 50 & 24.7 & 9.6 & 22.8 & \checkmark & \checkmarkcut & GW151226 & 153 & 12.3 & 11.1 & 5.3 & \checkmark & -- \\
GW170104 & 60 & 13.4 & 7.9 & 11.3 & \checkmark & \checkmarkcut & GW170608 & 179 & 15.8 & 14.8 & 6.3 & \checkmark & \checkmarkcut \\
GW170809 & 54 & 12.0 & 5.8 & 10.9 & -- & \checkmarkcut & GW170814 & 58 & 16.3 & 9.1 & 13.6 & \checkmark & \checkmarkcut \\
GW170818 & 48 & 10.8 & 4.5 & 10.1 & -- & \checkmarkcut & GW170823 & 40 & 11.5 & 4.2 & 11.1 & -- & \checkmarkcut \\
\midrule
\multicolumn{14}{c}{GWTC-2} \\
\hline
GW190408\_181802 & 68 & 15.0 & 8.3 & 12.5 & \checkmark & \checkmarkcut & GW190412 & 43 & 18.9 & 15.1 & 11.8 & \checkmark & \checkmark \\
GW190421\_213856 & 36 & 10.4 & 2.9 & 10.0 & -- & \checkmarkcut & GW190503\_185404 & 39 & 13.7 & 4.3 & 13.0 & -- & \checkmarkcut \\
GW190512\_180714 & 87 & 12.8 & 10.5 & 7.4 & \checkmark & \checkmarkcut & GW190513\_205428 & 48 & 13.3 & 5.1 & 12.2 & -- & \checkmarkcut \\
GW190517\_055101 & 41 & 11.1 & 3.4 & 10.5 & -- & \checkmarkcut & GW190519\_153544 & 23 & 15.0 & 0.0 & 15.0 & -- & \checkmarkcut \\
GW190521 & 14 & 13.9 & 0.0 & 13.9 & -- & \checkmarkcut & GW190521\_074359 & 40 & 25.4 & 9.7 & 23.5 & \checkmark & \checkmarkcut \\
GW190602\_175927 & 22 & 13.1 & 0.0 & 13.1 & -- & \checkmarkcut & GW190630\_185205 & 50 & 16.3 & 8.1 & 14.1 & \checkmark & \checkmarkcut \\
GW190706\_222641 & 19 & 12.7 & 0.0 & 12.7 & -- & \checkmarkcut & GW190707\_093326 & 161 & 13.4 & 12.2 & 5.5 & \checkmark & -- \\
GW190708\_232457 & 103 & 13.7 & 11.1 & 8.0 & \checkmark & \checkmarkcut & GW190720\_000836 & 126 & 10.5 & 9.2 & 5.2 & \checkmark & -- \\
GW190727\_060333 & 35 & 12.3 & 2.0 & 12.2 & -- & \checkmarkcut & GW190728\_064510 & 157 & 12.6 & 11.4 & 5.3 & \checkmark & -- \\
GW190814 & 147 & 24.6 & 22.9 & 9.1 & \checkmark & \checkmark & GW190828\_063405 & 45 & 16.2 & 6.0 & 15.1 & \checkmark & \checkmarkcut \\
GW190828\_065509 & 80 & 9.9 & 6.3 & 7.6 & \checkmark & \checkmarkcut & GW190910\_112807 & 35 & 14.4 & 3.3 & 14.0 & -- & \checkmarkcut \\
GW190915\_235702 & 46 & 13.1 & 3.7 & 12.6 & -- & \checkmarkcut & GW190924\_021846 & 239 & 12.2 & 11.8 & 3.4 & \checkmark & -- \\
\midrule
\multicolumn{14}{c}{GWTC-3} \\
\hline
GW191109\_010717 & 26 & 16.1 & 1.2 & 16.0 & -- & \starcutcheckmark & GW191129\_134029 & 222 & 12.5 & 11.9 & 3.9 & \checkmark & -- \\
GW191204\_171526 & 210 & 16.6 & 15.5 & 5.9 & \checkmark & -- & GW191215\_223052 & 73 & 10.5 & 6.1 & 8.4 & \checkmark & \checkmark \\
GW191216\_213338 & 213 & 18.0 & 16.8 & 6.4 & \checkmark & \checkmark & GW191222\_033537 & 36 & 11.8 & 2.8 & 11.5 & -- & \checkmark \\
GW200115\_042309 & 487 & 10.7 & 10.7 & 1.0 & \checkmark & -- & GW200129\_065458 & 62 & 25.8 & 12.3 & 22.6 & \checkmark & \checkmark \\
GW200202\_154313 & 233 & 10.3 & 9.8 & 3.0 & \checkmark & -- & GW200208\_130117 & 45 & 9.8 & 3.6 & 9.1 & -- & \checkmark \\
GW200219\_094415 & 40 & 9.5 & 2.3 & 9.2 & -- & \checkmark & GW200224\_222234 & 49 & 18.4 & 6.0 & 17.4 & -- & \checkmark \\
GW200225\_060421 & 98 & 11.8 & 8.0 & 8.7 & \checkmark & \checkmark & GW200311\_115853 & 56 & 16.5 & 6.9 & 15.0 & \checkmark & \checkmark \\
GW200316\_215756 & 182 & 9.8 & 9.0 & 3.7 & \checkmark & -- \\
\bottomrule[1pt]
\bottomrule[1pt]
\end{tabular}
\caption{List of events selected for the parameterized test that satisfy the criterion of FAR $< 10^{-3} \, \mathrm{yr}^{-1}$. The quantity $f_c^{\mathrm{PAR}}$  represents the cutoff frequency that separates the inspiral and postinspiral regimes, which is used for calculating the optimal SNRs in those two regimes. For the events in GWTC-1 and GWTC-2, this separation frequency corresponds to the transition frequency from inspiral to intermediate regime is given by the condition $GMf/c^3 = 0.018$, where $M$ is the total mass of the binary in the detector frame. In GWTC-3 events, this separation frequency corresponds to the dominant GW frequency at MECO, the orbit at which the orbital energy is at its lowest value. The quantities $\rho_{\mathrm{IMR}}$, $\rho_{\mathrm{insp}}$, and $\rho_{\mathrm{postinsp}}$ are the optimal SNRs of the full signal, the inspiral, and postinspiral regions respectively. All these listed values are the median of posterior samples. The last two columns specify whether the event is included in parametrized tests for the inspiral (PI), postinspiral (PPI), or both regimes. The listed values for GWTC-1 and GWTC-2 events were obtained from the public data provided by LVK collaboration~\cite{LIGOScientific:2019fpa, LIGOScientific:2020tif}. The symbol \checkmarkcut denotes that the corresponding event was used in previous LVK analyses, but we excluded it from this work. The symbol \starcutcheckmark denotes events that meet the selection criteria but are excluded from combined analysis to avoid false deviation due to data-quality issues. }
\label{tab:event_selection}
\end{table*}

\section{Analysis of GWTC-3 events}
\label{sec:GWTC-3}

\subsection{Event selection} 
A deviation parameter in the inspiral regime might not be well measurable, and the resulting posterior could be uninformative if the observed inspiral SNR of the signal is low and vice-versa for the post-inspiral parameters. Therefore, we analyze only those events that meet specific selection criteria. Following the event selection criteria considered in the previous analyses by the LVK collaboration~\cite{LIGOScientific:2019fpa, LIGOScientific:2020tif}, we consider the events that meet the significance threshold of  $\mathrm{FAR}< 10^{-3} \, \mathrm{yr}^{-1}$ and impose an additional requirement that the $\mathrm{SNR}>6$ in the inspiral regime. This criterion determines whether an event is included in the analyses with inspiral deviation parameters. Similarly, we apply the same SNR criterion in the post-inspiral regime to decide if an event qualifies for the analysis with post-inspiral parameters. As prescribed in the treatment of \phXAS model, we apply the \mbox{(2, 2)} mode frequency at MECO to divide the waveform between inspiral and post-inspiral regimes. Within the standard PN framework, the description of MECO with arbitrary mass ratio and spins is ill-defined. A previous study proposed a hybrid MECO approach to alleviate this issue by combining the information of PN theory with the exact Kerr solution~\citep{Cabero:2016ayq}. The corresponding function is implemented in \texttt{LAL}~\cite{lalsuite} and is labeled as \texttt{XLALSimIMRPhenomXfMECO}. We use that function to compute the cutoff frequency ($f_c^{\mathrm{PAR}}$) between inspiral and post-inspiral regimes.

In Table~\ref{tab:event_selection}, we present the optimal SNRs and cutoff frequencies for all selected events. The values listed for GWTC-1 and GWTC-2 events were obtained from analyses conducted by the LVK collaboration~\cite{LIGOScientific:2020tif}. Parametrized tests for these events were performed using the \phP model, where the cutoff frequency $f_c^{\mathrm{PAR}}$ is determined by the condition $GMf/c^3 = 0.018$, with $M$ representing the binary’s total mass in the detector frame.

For GWTC-3 events, we used public posterior samples with the tag \texttt{C01:IMRPhenomXPHM}, provided by the LVK collaboration and accessible on Zenodo~\cite{ligo_scientific_collaboration_and_virgo_2023_8177023}. The bottom section of Table~\ref{tab:event_selection} reports the median values of cutoff frequencies and SNRs for GWTC-3 events. We exclude the event GW191109\_010717 from the combined analysis, even though it satisfies the event selection criteria for post-inspiral analysis. We found apparent deviations for this event, similar to those reported in the tests of GR paper with GWTC-3 by LVK~\cite{LIGOScientific:2021sio}. Data quality issues in both LIGO detectors are considered to be the explanation for the apparent tension, rather than genuine departures from GR. Out of the 15 events, 10 meet the criteria for the analysis of inspiral parameters, and another 9 for post-inspiral parameters.

Two notable events, GW190412 and GW190814, were separately highlighted in the GWTC-2 testing GR analyses by the LVK due to the strong presence of higher-order modes and precession in their signals~\cite{LIGOScientific:2020tif}. We also include these two events in our analysis alongside the GWTC-3 events. Both events satisfy the selection criteria for both inspiral and post-inspiral deviation parameters. To evaluate the event selection criteria, we use the \phXPHM posterior samples provided in the GWTC-2.1 catalog by the LVK~\cite{LIGOScientific:2021usb}.

\begin{figure*}
    \centering
    \includegraphics[width=0.95\textwidth]{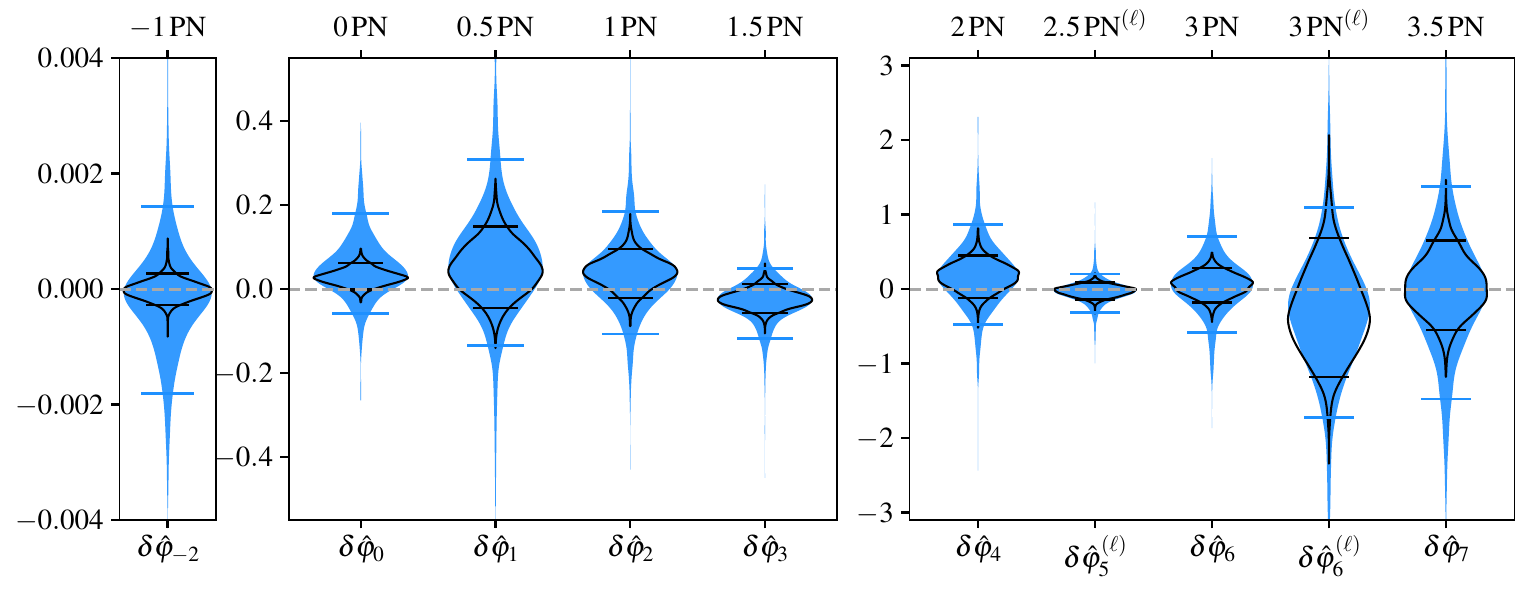}
    \caption{Illustrating the combined results for the inspiral deviation parameters, obtained from the GWTC-1/2/3 events listed in Table~\ref{tab:event_selection}. The filled color probability distributions represent the results from the hierarchical approach, while the unfilled black solid lines show the combined results assuming a common value for each deviation parameter across all events. In contrast, the hierarchical approach allows for an independent value for each event. The horizontal blue ticks indicate the 90\% credible intervals derived from the hierarchical analyses.}
    \label{fig:violin_gwtc3_insp}
\end{figure*}
\begin{figure*}
    \centering
    \includegraphics[width=0.95\textwidth]{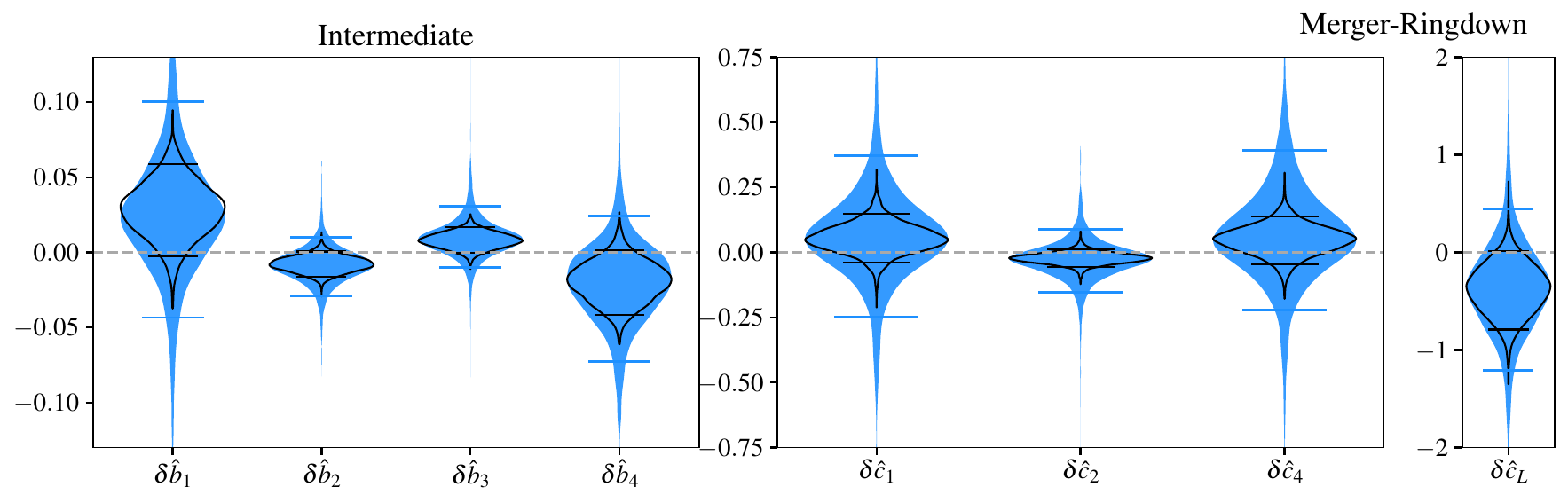}
    \caption{Same as Fig.~\ref{fig:violin_gwtc3_insp} but showing the combined results for the post-inspiral deviation parameters, obtained from the GW190412, GW190814, and GWTC-3 events listed in Table~\ref{tab:event_selection}.}
    \label{fig:violin_gwtc3_postinsp}
\end{figure*}

\begin{figure*}[!t]
    \centering
    \includegraphics[width=0.95\textwidth]{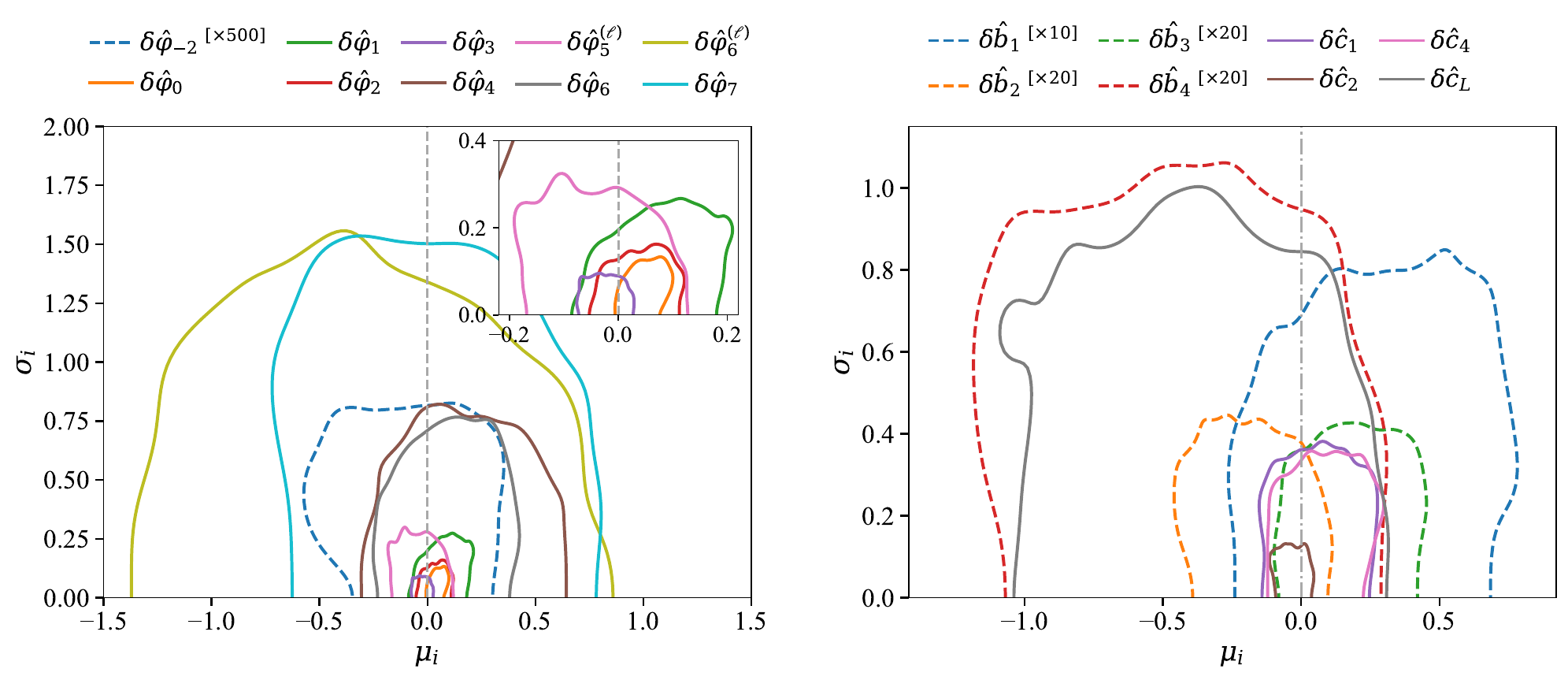}
    \caption{Illustrating the joint distribution for the hyperparameters $\mu_i$ and $\sigma_i$ of the beyond-GR parameters $\dpi_i$ for the parametrized tests. The left panel shows the results for inspiral parameters obtained using the selected events in GWTC-1/2/3 as reported in Table~\ref{tab:event_selection}. The right panel shows the results for post-inspiral parameters obtained using the selected events in GWTC-3, GW190412, and GW190814. The contours denote the 90\% credible regions of the joint distribution. The contours for $\delta\hat{\varphi}_{-2}$ and intermediate parameters ($\delta\hat{b}_i$) are rescaled by a factor of 1000 and 20, respectively, to improve visibility. All contours include $\mu_i = \sigma_i = 0$, consistent with the GR prediction.}
    \label{fig:par_hier_contour}
\end{figure*}
\begin{table*}[!t]
\centering
\setlength{\tabcolsep}{2pt}
\renewcommand{\arraystretch}{1.6}
\begin{tabular}{ l c c c c c c @{\hskip 0.3cm} || @{\hskip 0.2cm} c c c c c c c }
\toprule[1pt]
\toprule[1pt]
$\hat{p}_i$ & \multicolumn{4}{c}{Hierarchical} & \multicolumn{2}{c||@{\hskip 0.2cm}}{Restricted} & $\hat{p}_i$ & \multicolumn{4}{c}{Hierarchical} & \multicolumn{2}{c}{Restricted} \\
\cmidrule[0.8pt](lr{0.75em}){2-5} \cmidrule[0.8pt](lr{0.75em}){6-7} \cmidrule[0.8pt](lr{0.75em}){9-12} \cmidrule[0.8pt](lr{0.75em}){13-14}
   & $\mu_i$ & $\sigma_i$ & $\dpi_i$ & $\QGR$[\%] & $\dpi_i$ & $\QGR$[\%] & 
   & $\mu_i$ & $\sigma_i$ & $\dpi_i$ & $\QGR$[\%] & $\dpi_i$ & $\QGR$[\%] \\
\midrule[0.8pt]
$\varphi_{-2}\substack{[ \times 500 ]}$  & $-0.06^{+0.34}_{-0.43}$ &   $< 0.72$  &   $-0.05^{+0.77}_{-0.86}$ &   16.1 &   $-0.01^{+0.15}_{-0.13}$ &   14.4 & $b_1\substack{[ \times 20 ]}$  & $0.49^{+0.82}_{-0.74}$ &   $< 1.32$  &   $0.48^{+1.52}_{-1.35}$ &   54.0 &   $0.57^{+0.6}_{-0.63}$ &   85.7 \\ 
$\varphi_{0}$  & $0.04^{+0.06}_{-0.04}$ &   $< 0.11$  &   $0.04^{+0.14}_{-0.1}$ &   42.3 &   $0.03^{+0.03}_{-0.03}$ &   84.4 & $b_2\substack{[ \times 20 ]}$  & $-0.18^{+0.22}_{-0.21}$ &   $< 0.35$  &   $-0.17^{+0.37}_{-0.41}$ &   55.1 &   $-0.16^{+0.17}_{-0.17}$ &   87.8 \\  
$\varphi_{1}$  & $0.07^{+0.12}_{-0.11}$ &   $< 0.21$  &   $0.06^{+0.25}_{-0.2}$ &   43.7 &   $0.05^{+0.1}_{-0.09}$ &   42.1 & $b_3\substack{[ \times 20 ]}$  & $0.19^{+0.22}_{-0.23}$ &   $< 0.35$  &   $0.18^{+0.44}_{-0.38}$ &   63.6 &   $0.16^{+0.17}_{-0.17}$ &   86.3 \\ 
$\varphi_{2}$  & $0.04^{+0.07}_{-0.07}$ &   $< 0.12$  &   $0.03^{+0.15}_{-0.14}$ &   37.3 &   $0.04^{+0.06}_{-0.06}$ &   71.8 & $b_4\substack{[ \times 20 ]}$  & $-0.41^{+0.52}_{-0.63}$ &   $< 0.83$  &   $-0.39^{+0.87}_{-1.06}$ &   34.3 &   $-0.38^{+0.42}_{-0.45}$ &   87.8 \\ 
$\varphi_{3}$  & $-0.03^{+0.04}_{-0.04}$ &   $< 0.08$  &   $-0.02^{+0.07}_{-0.09}$ &   47.5 &   $-0.02^{+0.04}_{-0.03}$ &   77.9 & $c_1$  & $0.06^{+0.17}_{-0.16}$ &   $< 0.3$  &   $0.06^{+0.31}_{-0.31}$ &   33.1 &   $0.05^{+0.1}_{-0.09}$ &   62.6 \\ 
$\varphi_{4}$  & $0.18^{+0.39}_{-0.37}$ &   $< 0.63$  &   $0.18^{+0.69}_{-0.66}$ &   35.5 &   $0.17^{+0.28}_{-0.29}$ &   69.7 & $c_2$  & $-0.03^{+0.06}_{-0.07}$ &   $< 0.11$  &   $-0.03^{+0.12}_{-0.13}$ &   31.2 &   $-0.02^{+0.04}_{-0.04}$ &   74.3 \\  
$\varphi_{5}^{(\ell)}$  & $-0.03^{+0.11}_{-0.13}$ &   $< 0.24$  &   $-0.03^{+0.23}_{-0.28}$ &   19.5 &   $-0.02^{+0.11}_{-0.12}$ &   6.0 &  $c_4$  & $0.07^{+0.17}_{-0.14}$ &   $< 0.28$  &   $0.06^{+0.33}_{-0.28}$ &   23.6 &   $0.05^{+0.09}_{-0.1}$ &   63.1 \\ 
$\varphi_{6}$  & $0.09^{+0.25}_{-0.27}$ &   $< 0.58$  &   $0.07^{+0.64}_{-0.65}$ &   12.0 &   $0.06^{+0.22}_{-0.24}$ &   30.0 &  $c_{L}$  & $-0.37^{+0.47}_{-0.49}$ &   $< 0.69$  &   $-0.36^{+0.81}_{-0.85}$ &   67.8 &   $-0.37^{+0.39}_{-0.42}$ &   88.0 \\
$\varphi_{6}^{(\ell)}$  & $-0.29^{+0.92}_{-0.87}$ &   $< 1.19$  &   $-0.28^{+1.38}_{-1.44}$ &   21.2 &   $-0.33^{+1.02}_{-0.85}$ &   53.6 \\ 
$\varphi_{7}$  & $0.01^{+0.64}_{-0.66}$ &   $< 1.28$  &   $-0.0^{+1.37}_{-1.47}$ &   7.2 &   $0.03^{+0.63}_{-0.58}$ &   16.9 \\
\bottomrule[1pt]
\bottomrule[1pt]
\end{tabular}
\caption{Summary of the combined results for each deviation parameter $\dpi_i$. All quantities represent the median and 90\%-credible intervals except $\sigma_i$, for which we provide an upper limit. For both general and restricted results, $\QGR$ is the GR quantile associated with Fig.~\ref{fig:violin_gwtc3_insp} and \ref{fig:violin_gwtc3_postinsp}.}
\label{tab:summaryGWTC3}
\end{table*}

\subsection{Combining multiple events}

In the context of testing GR with multiple events, we typically summarize the findings with a single statement of agreement with GR by combining the results from individual events. This approach can enhance the detectability of potential deviations or constrain them more tightly. LVK analyses usually consider two approaches: (a) multiplying the posterior distributions from all events, assuming a common beyond-GR parameter across them, and (b) employing a hierarchical approach, which assumes that these parameters follow an underlying distribution.

\subsubsection{Common parameter and multiplying posteriors}
If we assume the deviation parameter value is the same across the events, we can combine the evidence to measure that parameter by multiplying individual posteriors for each event. For $N$ events, lets assume the posterior distribution $\displaystyle p(\dpi^{(j)}_i \mid d^{(j)})$ obtained from the $j$-th data $d^{(j)}$ for the deviation parameter $\dpi_i$, where $j=1,...,N$. The joint posterior probability is given as, 
\begin{align}
\displaystyle p\left(\dpi_i \mid \{ d^{(j)} \}_{j=1}^N \right) & \propto \prod_{j=1}^{N} p\left(\dpi_i \mid d^{(j)}\right)  \\ 
&  \propto \prod_{j=1}^{N} p\left(d^{(j)} \mid \dpi_i \right) \; p\left(\dpi_i\right),
\end{align}
where $p\big(d^{(j)} \mid \dpi_i\big)$ is the likelihood and $p(\dpi_i)$ is prior. In the parametrized test, we generally consider uniform prior distribution for $\dpi_i$. In that case, the joint posterior is proportional to the product of the individual likelihoods and a single instance of the prior~\cite{Meidam:2017dgf, Zimmerman:2019wzo}. However, in practice, we obtain the marginalized posterior samples of a deviation parameter for each event. We use Gaussian KDE to approximate the posterior distribution for each set of samples. The individual KDEs are then multiplied pointwise to estimate the combined posterior. We evaluate all KDEs on a common grid using the \texttt{gaussian\_kde} code in \scipy~\cite{Virtanen:2019joe}.

\subsubsection{Hierarchical approach}
Combining multiple events, assuming the deviation parameter is constant across the events, could lead to an incorrect conclusion. %This is justified for studies of modified GW propagation, since those effects should not depend on the source. In other analyses, 
We expect their value to vary with different source properties and choice of alternative theory. As described in Refs~\cite{Isi:2019asy, Zimmerman:2019wzo}, we can relax the assumption of a common parameter across all events by using the hierarchical approach. Instead of a common parameter, this method assumes the non-GR parameters are drawn from a common underlying distribution, and then estimates the properties of that distribution using a population of events. Following the previous studies, we model the population distribution of non-GR parameters with a Gaussian distribution, $\dpi_i \sim \mathcal{N}(\mu_i, \sigma_i)$, where $\mu_i$ and $\sigma_i$ are unknown mean and standard deviation, respectively. Setting $\sigma_i=0$ amounts to assuming that all systems share the same beyond-GR parameter $\dpi_i$. From the Bayes theorem, the joint posterior on $\mu_i$ and $\sigma_i$ is given as,
\begin{align}
\label{eq:hyper_posterior}
\displaystyle p\left(\mu_i, \sigma_i \mid \{ d^{(j)} \}_{j=1}^N \right) & \propto \prod_{j=1}^{N} p\left(d^{(j)} \mid \mu_i, \sigma_i \right) \; p(\mu_i, \sigma_i)  \nonumber \\ 
&  \propto p(\mu_i, \sigma_i) \prod_{j=1}^{N} p\left(d^{(j)} \mid \mu_i, \sigma_i \right),
\end{align}
where $p(\mu_i, \sigma_i)$ is the prior on hyperparameters $(\mu_i, \sigma_i)$. We assume these to be uniformly distributed and the same for all events. The quantity $p\big(d^{(j)} \mid \mu_i, \sigma_i \big)$ is individual likelihood for the deviation parameter $\dpi_i$,
\begin{equation}
\label{eq:hier_likelihood}
p\big(d^{(j)} \mid \mu_i, \sigma_i \big) = \int p\left(d^{(j)} \mid \dpi_i\right) \: p\big(\dpi_i \mid \mu_i, \sigma_i\big) \: d\big[\dpi_i\big],
\end{equation}
where $p\big(\dpi_i \mid \mu_i, \sigma_i\big) = \mathcal{N}(\mu_i, \sigma_i)$ by construction. The quantity $p\big(d^{(j)} \mid \dpi_i\big)$ is the standard likelihood obtained from the standard parameter estimation analyses. In practice, we build a Gaussian KDE using the marginalized posterior samples of the deviation parameter as given in Eq.~\eqref{eq:dpi_posterior},  which simplifies the evaluation of the integral in  Eq.~\eqref{eq:hier_likelihood}. After that, we evaluate the joint posterior on hyperparameters in Eq.~\eqref{eq:hyper_posterior} using a Hamiltonian Monte Carlo algorithm implemented in the \texttt{stan} package~\cite{carpenter2017stan}. Finally, using the posterior of hyperparameters $p\big(\mu_i, \sigma_i \mid \{ d^{(j)} \}_{j=1}^N \big)$, we infer the combined posterior of $\dpi_i$ by marginalizing over $\mu_i$ and $\sigma_i$,
\begin{align}
p\left(\dpi_i \mid \{ d^{(j)} \}_{j=1}^N \right) = \int & p\big(\dpi_i \mid \mu_i, \sigma_i\big) \nonumber \\ 
& \times p\left(\mu_i, \sigma_i \mid \{ d^{(j)} \}_{j=1}^N \right) \; d\mu_i \: d\sigma_i
\end{align}
If GR is correct, then both hyperparameters, $\mu_i$ and $\sigma_i$, are expected to be consistent with zero. If we find a nonzero $\mu_i$, this is an obvious deviation from GR or a systematic error in the analysis of one or several of the events under consideration.

\subsection{Results and Discussion}

The unfilled black solid violin plot in Fig.~\ref{fig:violin_gwtc3_insp} and Fig.~\ref{fig:violin_gwtc3_postinsp} shows the combined results obtained using the common parameter approach. Results with the hierarchical approach are shown in filled blue distributions. In both cases, all the distributions are consistent with GR, $\dpi_i=0$, and always within 90\% credible symmetric interval. Fig.~\ref{fig:par_hier_contour} shows the 90\% credible region of the joint distribution for the hyperparameters $\mu$ and $\sigma$. We find a population that the population of all beyond-GR parameters is consistent with GR for both in $\mu_i$ and $\sigma_i$. All $\mu_i$ posteriors are consistent with 0 at the $0.5\sigma$ level or better, while all $\sigma_i$ posteriors peak at 0.

\begin{figure*}[!t]
    \centering
    \includegraphics[width=0.95\textwidth]{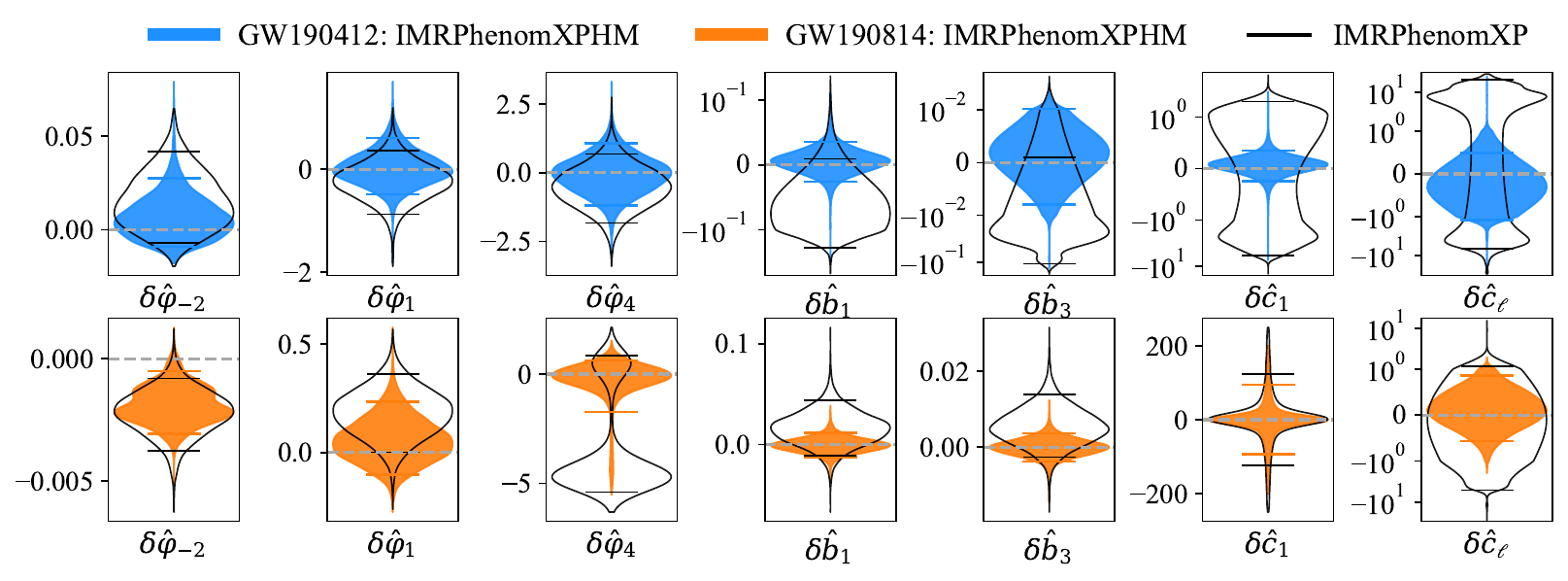}
    \caption{Illustrating the posterior distributions of parametrized deviation parameters obtained from GW190412 (shown in blue) and GW190814 (shown in orange) events, inferred using \phXPHM model. %The black unfilled violin plots in top (GW190412) and middle (GW190814) panels show the results obtained from the GWTC-2 analyses by LVK, as shown in Fig.\:19 of Ref.\:\cite{LIGOScientific:2020tif}.
    The black unfilled violin plots show the results inferred using \phXP model, highlighting the systematic biases when higher-order modes are missing in the TIGER baseline waveform model. The horizontal solid lines indicate the 90\% credible intervals, and the gray dashed line at $\dpi_i = 0$ denotes the GR values. The full results are shown in Fig.~\ref{fig:violin_gw190412_gw190814_full}.}
    \label{fig:violin_gw190412_gw190814}
\end{figure*}

We summarize the combined results in Table~\ref{tab:summaryGWTC3}. The hierarchical analysis yields the tightest bound with the -1PN term, \mbox{$\dphi_{-2}=-0.01^{+0.15}_{-0.13} \times 0.002$}, within a 90\% credible interval. The next tightest bounds are found with intermediate parameters, particularly $\delta\hat{b}_3=0.16^{+0.17}_{-0.17}\times 0.05$, within the same interval. The widest bound appears with the 3PN log term, $\dphi_{6}^{(\ell)}= -0.33^{+1.02}_{-0.85}$. In Table~\ref{tab:summaryGWTC3}, we also report the results from the common parameter analysis under the “Restricted” column, showing consistently tighter 90\% credible intervals for $\dpi_i$ compared to the hierarchical analysis. This is expected as common parameter analysis is a special case of hierarchical analysis works assuming $\sigma_i=0$, which can only explore a subspace of entire population of the non-GR parameters. We note that this can lead to biased conclusion if that assumption is not correct. While comparing these two approaches for combining the events, we find significant changes in 0PN, 3.5PN, and several post-inspiral parameters, as reported in terms of \( \QGR \) in Table~\ref{tab:summaryGWTC3}.

% in terms of $\QGR$ as given in Eq.~\eqref{eq:qgr}.

Two widely used approaches in the context of parametrized tests of GR with GW signals are TIGER and FTI. In Appendix~\ref{appendix:TIGERvsFTI}, we compare TIGER results against FTI results reported by LVK. We highlight the results for two O3b events, GW200225\_060421 and GW191204\_171526, where the first one shows excellent agreement, and the other has inconsistency in many of the PN terms. The baseline models of these approaches are not the same, and the FTI model neglects the contribution of higher-order modes and the precession effect, which could lead to disagreement. For GWTC-3 events, we have not noticed any significant inconsistency. The combined results reported in Fig.~\ref{fig:violin_gwtc3_insp} are also comparable to FTI, except for the -1PN parameter.
%The improvement in -1PN for TIGER results is due to the two GWTC-2 events, GW190814 and GW190412, where the contribution of higher-order modes leads to a narrower posterior, as shown in Fig.~\ref{fig:tiger_vs_fti}.

\subsection{Analysis of GW190412 and GW190814}
\label{sec:gw190412_gw190814}
Two notable events in GWTC-2, GW190412 and GW190814, originated from compact binary mergers with highly asymmetric component masses, which enabled the first observation of gravitational-wave radiation beyond the $(\ell, m) = (2, 2)$ mode~\cite{LIGOScientific:2020stg, LIGOScientific:2020zkf, Roy:2019phx}. The presence of higher harmonics has enabled many new tests of GR by examining how these higher-order modes are related to the primary signal~\cite{Kastha:2018bcr, Mahapatra:2023ydi, Dhanpal:2018ufk, Puecher:2022sfm, Islam:2019dmk, Capano:2020dix, Mahapatra:2023hqq, Mahapatra:2023uwd}. Measuring these higher harmonics of the signal also provides a more accurate measurement of source parameters, which allows for tighter constraints on certain deviations from GR. 

To assess the improvements due to higher harmonics, we perform the TIGER analysis using two different waveform models: \phXP and \phXPHM. Fig.~\ref{fig:violin_gw190412_gw190814} highlights the biases and improvements in the posterior of deviation parameters when higher harmonics are included in the model compared to a model without them. The filled violins represent the posterior for the \phXPHM model, while the black unfilled violins correspond to the \phXP model. Fig.~\ref{fig:violin_gw190412_gw190814} show the results for a part of the deviation parameters, violin plots for all the deviation parameters are shown in Fig.~\ref{fig:violin_gw190412_gw190814_full}.

For both the events, the posteriors with \phXPHM model are more consistent with GR value compared to the case of \phXP model. We observe noticeable improvement in the post-inspiral parameters. 
Another noticeable improvement is seen in 2PN term for GW190814 event. The $\dphi_4$ posterior exhibits bimodality, with the stronger peak deviating from the GR value. This behavior is associated with the bimodality in the mass ratio and component masses. This significant improvement indicates the crucial role of higher harmonics in modeling TIGER waveforms to achieve unbiased conclusions in testing GR analysis.

We note that in the analyses with GW190814, the GR value lies outside the 90\% credible interval for several testing parameters (-1PN, 3PN, and 3.5PN). This is not necessarily surprising, as the GR value is found near the tail of the distribution. A similar trend was also observed in earlier analyses reported by the LVK in Fig. 19 of Ref.~\cite{LIGOScientific:2020tif}. However, for the posterior obtained using the FTI framework, the GR value lies within the 90\% credible interval, albeit near the edge, and a partially bimodal shape is also observed in several testing parameters, as shown in the bottom row of Fig.~\ref{fig:tiger_vs_fti}. We further note that the 0PN posterior for the GW230529 event was found to deviate significantly from the GR value~\cite{Sanger:2024axs}. This is mostly due to strong degeneracies between the testing parameter and the physical parameters of the binary in certain regions of the parameter space of the parametrized waveform.

\section{Summary and conclusion}
\label{sec:conclusion}

We have developed a parameterized testing framework based on the \phX models, incorporating spin precession and higher-order modes. Parameterized deviations are introduced in the phasing coefficients of the dominant quadrupole mode \mbox{$(\ell, m) = (2, 2)$} for non-precessing binaries. This new implementation allows deviations to propagate into the higher-order modes during the inspiral phase. When generating parameterized waveforms with precessing spin models like \phXP and \phXPHM, the non-GR correction undergoes a twist-up due to the transformation of waveform modes from the co-precessing frame to the inertial frame. However, our beyond-GR correction does not interfere with the determination of spin dynamics.

This TIGER framework with \phX models is integrated into the LIGO Algorithm Library (LAL)~\cite{lalsuite}, the parametrized waveforms can be generated using the \lalsimulation library~\footnote{Available at \href{https://git.ligo.org/lscsoft/lalsuite}{https://git.ligo.org/lscsoft/lalsuite}. There was issue in the -1PN implementation, }. The corresponding Bayesian analysis framework is implemented in \bilbytgr package~\cite{bilby-tgr}. While compiling this manuscript, this TIGER framework had already been applied in various studies, including testing GR with the GW230529 event~\cite{Sanger:2024axs, LIGOScientific:2024elc} and investigating astrophysical environmental effects in tests of vacuum GR~\cite{Roy:2024rhe}.

With this framework, we conducted a series of injection studies on GW200129-like events to investigate the impact of spin precession and higher-order modes if either is omitted in the baseline TIGER waveform, as discussed in Sec.~\ref{sec:systematic}. Our results indicate that if these contributions are significant, neglecting one of them can lead to biased conclusions, suggesting false deviations from GR. We also noticed systematic biases in most deviation parameters, even when these contributions are relatively small. Previous studies on systematic biases also reached to the similar conclusions using the earlier version of TIGER~\cite{Pang:2018hjb}, where higher-order modes were not included in the baseline model. Our new implementation resolves these systematic biases.

We then performed the analyses for the events in GWTC-3 and reported a combined statement by including the events in GWTC-1/2. We also performed the analyses for two notable events GW190412 and GW190814, which were associated with highly asymmetric binary and in which strong contribution of higher-order modes were observed. To provide the combined statement, we made use of two techniques: common parameter approach and hierarchical approach. With both approaches, we found the combined posterior is consistent with the prediction of GR. 

We note that while the \phX waveform family has undergone several recent improvements~\cite{Hamilton:2021pkf, Thompson:2023ase, Colleoni:2024knd, Hamilton:2025xru}, the TIGER results presented in this paper were obtained using \phXPHMMSA (the \phXPHM model with the multi–scale analysis (MSA) treatment of spin precession)~\cite{Pratten:2020ceb}, consistent with the waveform choices in the GWTC-3 catalog~\cite{KAGRA:2021vkt}. A recent upgrade, \phXPHMST~\cite{Colleoni:2024knd}, is now being employed for parameter estimation of LVK events observed during the fourth observing run. This version uses the \textsc{SpinTaylor} framework~\cite{Sturani:2019} to numerically evolve the spins, providing a more accurate precession evolution compared to the analytical MSA approximation. The TIGER framework works straightforwardly with \phXPHMST, since this model employ the same underlying co-precessing \phXHM implementation as in \phXPHMMSA. Further efforts were made to improve the spin-precession sector of the \phX waveform family by calibrating the precession angles to numerical relativity, leading to the waveform models \phXOfoura~\cite{Hamilton:2021pkf, Thompson:2023ase} and \phXPNR~\cite{Hamilton:2025xru} that have more recently become available. However, those introduce introduce additional modifications to the co-precessing frame, requiring a dedicated TIGER implementation and further validation.

\section*{Acknowledgements}
We are grateful to Michalis Agathos, Justin Janquart and Elise M. S\"anger for useful comments on our draft. S.R. is supported by the Fonds de la Recherche Scientifique - FNRS (Belgium). G.P. gratefully acknowledges support from a Royal Society University Research Fellowship URF{\textbackslash}R1{\textbackslash}221500 and RF{\textbackslash}ERE{\textbackslash}221015, as well as STFC grants ST/V005677/1 and ST/Y00423X/1.
M.H., P.T.H.P., and C.V.D.B. are supported by the research programme of the Netherlands Organisation for Scientific Research (NWO).
The authors are grateful for computational resources provided by the LIGO Laboratory and supported by National Science Foundation Grants PHY-0757058 and PHY-0823459. The authors are also grateful for the computational resources provided by Cardiff University supported by STFC grant ST/I006285/1.
This research has made use of data or software obtained from the Gravitational Wave Open Science Center (\href{https://www.gwosc.org}{https://www.gwosc.org}), a service of the LIGO Scientific Collaboration, the Virgo Collaboration, and KAGRA. This material is based upon work supported by NSF's LIGO Laboratory which is a major facility fully funded by the National Science Foundation, as well as the Science and Technology Facilities Council (STFC) of the United Kingdom, the Max-Planck-Society (MPS), and the State of Niedersachsen/Germany for support of the construction of Advanced LIGO and construction and operation of the GEO600 detector. Additional support for Advanced LIGO was provided by the Australian Research Council. Virgo is funded, through the European Gravitational Observatory (EGO), by the French Centre National de Recherche Scientifique (CNRS), the Italian Istituto Nazionale di Fisica Nucleare (INFN) and the Dutch Nikhef, with contributions by institutions from Belgium, Germany, Greece, Hungary, Ireland, Japan, Monaco, Poland, Portugal, Spain. KAGRA is supported by Ministry of Education, Culture, Sports, Science and Technology (MEXT), Japan Society for the Promotion of Science (JSPS) in Japan; National Research Foundation (NRF) and Ministry of Science and ICT (MSIT) in Korea; Academia Sinica (AS) and National Science and Technology Council (NSTC) in Taiwan. We have used \numpy~\cite{Harris:2020xlr} and \scipy~\cite{Virtanen:2019joe} for analyses, and \matplotlib~\cite{Hunter:2007ouj} for preparing the plots in this manuscript.

\appendix
\section{Impact of higher harmonics and precession}
\label{sec:appendixA}
\begin{figure*}
    \centering
    \includegraphics[width=0.98\textwidth]{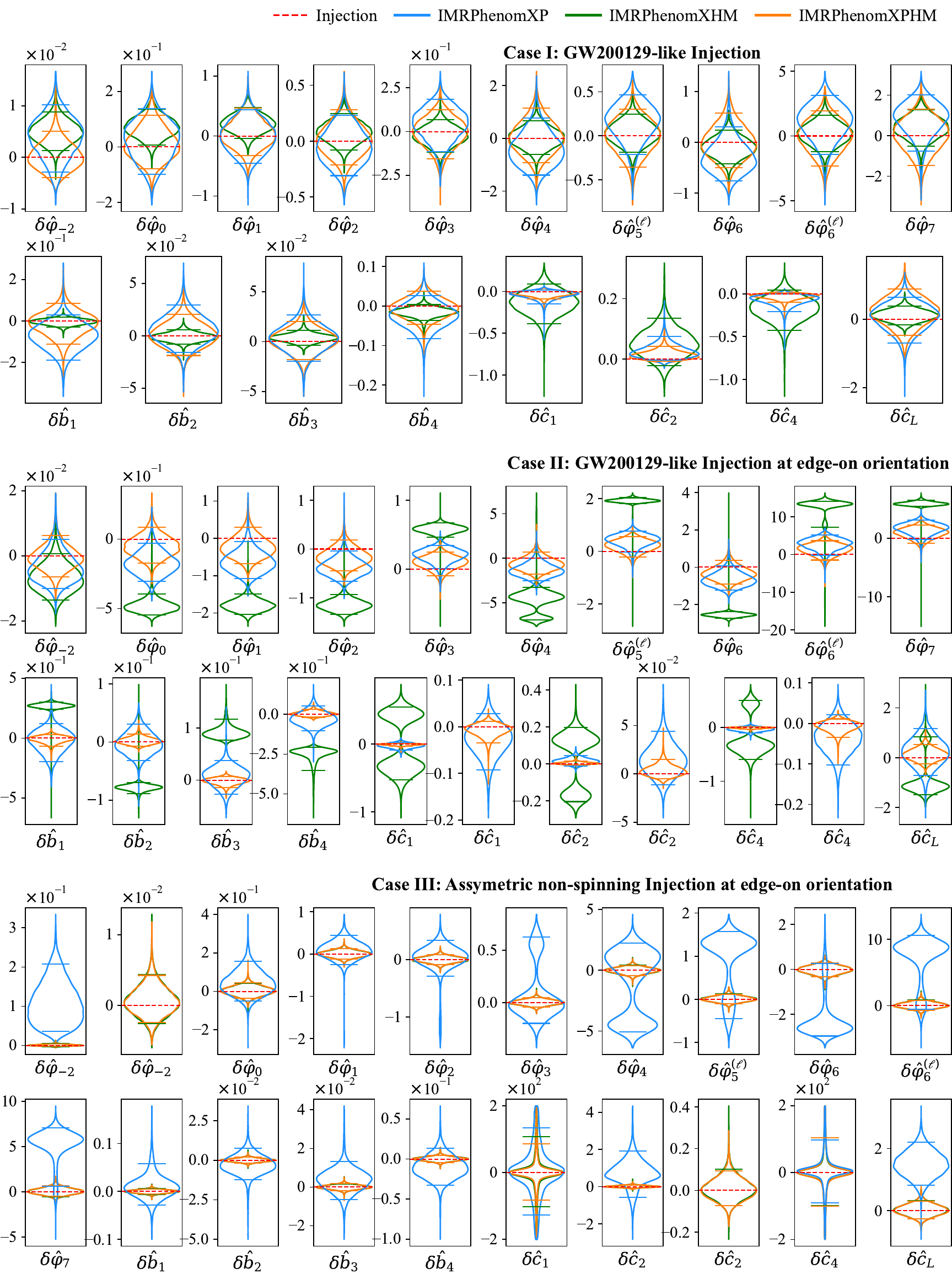}
    \caption{Same as Fig.\:\ref{fig:gw200129-like_inj}, but displaying the posterior distributions for all deviation parameters. For some deviation parameters, the posteriors obtained using \phXP are significantly broader than those from the other two models. These cases are shown in two consecutive subplots to highlight the difference and to clearly display the narrower posteriors. These TIGER results for the GW200129-like injections are obtained using different waveform models to highlight the impact of higher-order modes and precession effects in parametrized tests, as described in Sec.~\ref{sec:systematic} and Appendix ~\ref{sec:appendixA}.}
    \label{fig:gw200129-like_inj_full}
\end{figure*}

\begin{figure*}[t]
    \centering
    \includegraphics[width=0.98\textwidth]{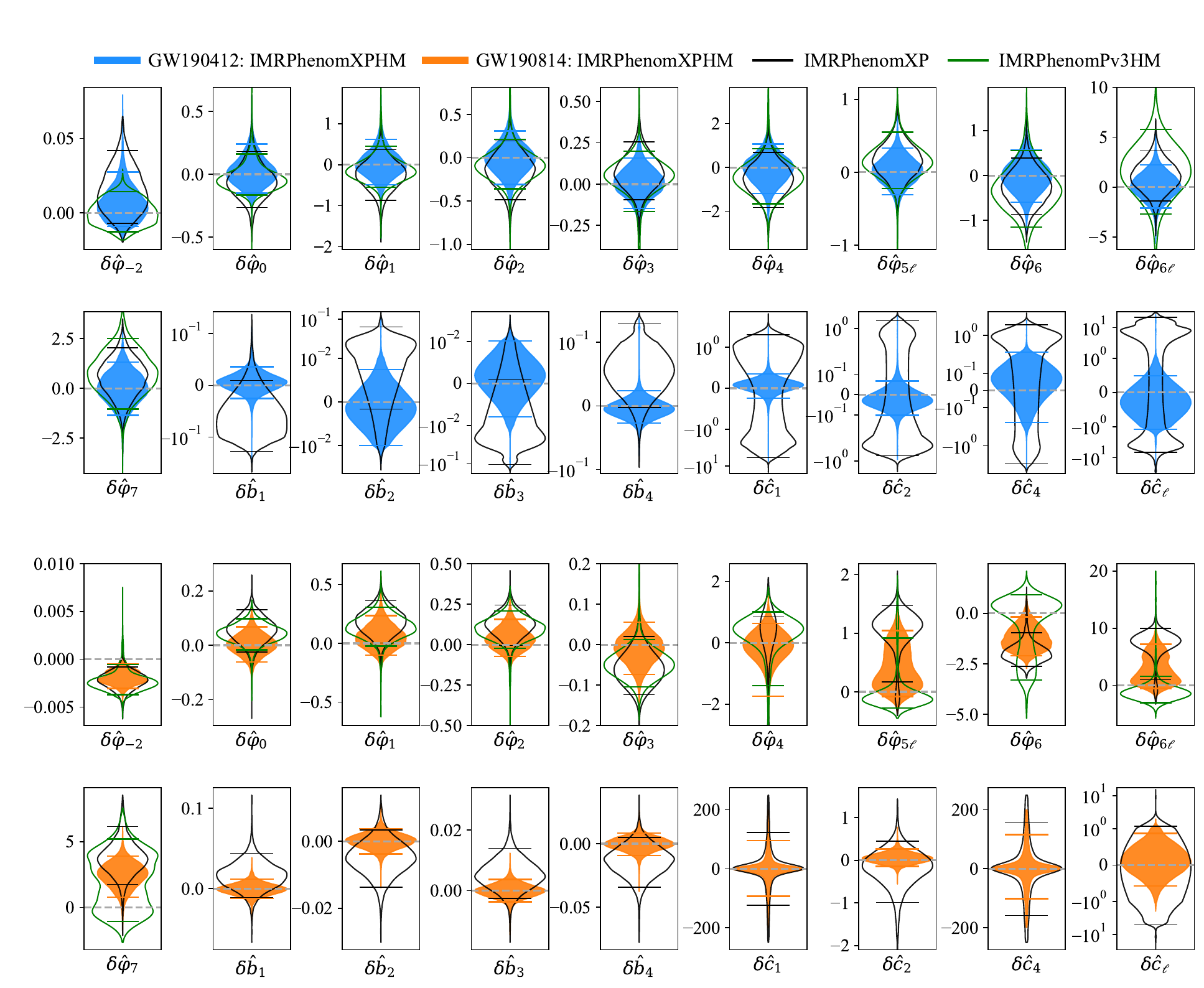}
    \caption{Same as Fig.~\ref{fig:violin_gw190412_gw190814}, but showing the posterior distributions for all deviation parameters. These TIGER results for the GW190412 and GW190814 events are obtained using different waveform models to highlight the impact of higher-order modes and precession effects in parametrized tests, as discussed in Sec.~\ref{sec:gw190412_gw190814} and Appendix ~\ref{sec:appendixA}. For the inspiral parameters, we also show the results obtained with the \phPvHM model, as reported in the GWTC-2 tests of GR paper by LVK~\cite{LIGOScientific:2020tif}.}
    \label{fig:violin_gw190412_gw190814_full}
\end{figure*}

In this section, we highlight systematic biases that can arise from neglecting higher-order modes and precession in the baseline waveform of the parametrized test, which can lead to a false indication of deviation from GR. Here, we show the posterior of all the deviation parameters implemented in the new TIGER framework. Fig.~\ref{fig:gw200129-like_inj_full} shows the results for GW200129-like injection described in Sec.~\ref{sec:systematic}. Fig.~\ref{fig:violin_gw190412_gw190814_full} shows the full results obtained for the GW190412 and GW190814 events, as demonstrated in Sec.~\ref{sec:gw190412_gw190814}.

\section{Comparison between TIGER and FTI}
\label{appendix:TIGERvsFTI}
Another common approach for parametrized tests of GR with GW observations is known as FTI~\cite{Mehta:2022pcn}. Similar to the TIGER method, FTI introduces deviations in the phasing coefficients to generate a non-GR waveform. However, instead of directly incorporating deviation parameters into the frequency-domain phase coefficients, it computes the non-GR correction for a given PN term and then adds it to the frequency-domain phase of the GR waveform. To ensure that the post-inspiral part of the waveform remains unaffected by the non-GR correction, a windowing function is applied to smoothly taper it off to zero. The current FTI framework is applicable to any aligned-spin waveform model. The FTI results in the LVK analysis were obtained using the \seobnrrom model~\cite{Bohe:2016gbl}, a frequency-domain waveform model for aligned-spin binaries that neglects higher-order modes and precession effects.

\begin{figure*}[!t]
    \centering
    \includegraphics[width=0.98\textwidth]{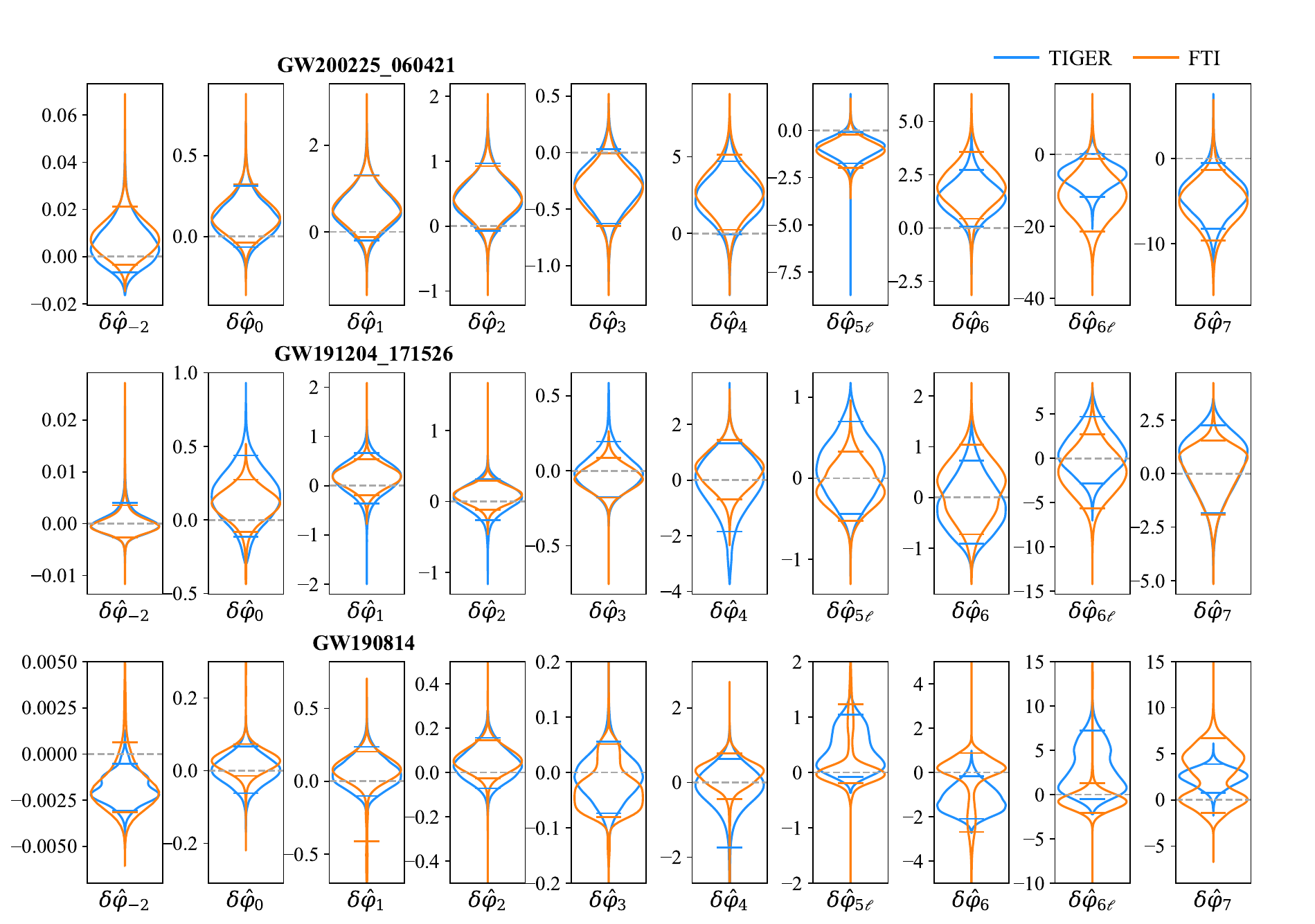}
    \caption{Comparison of TIGER and FTI results. The top and middle rows correspond to the O3b events GW200225\_060421 and GW191204\_171526, respectively. The bottom row shows the comparison for the GW190814 event. The FTI results are obtained from the LVK data release~\cite{GWTC3:TGR:release}. These results are further discussed in Appendix~\ref{appendix:TIGERvsFTI}. The small horizontal lines indicate the 90\% credible intervals of the posteriors, and the gray dashed line at $\dphi_i=0$ represents the GR prediction.}
    \label{fig:tiger_vs_fti}
\end{figure*}

We present a comparison between our TIGER results and FTI results reported in Ref.~\cite{LIGOScientific:2021sio}. The data for the FTI analysis has been obtained from the LVK data release~\cite{GWTC3:TGR:release}. Fig.~\ref{fig:tiger_vs_fti} highlights the comparison for two O3b events. For the GW200225\_060421 event, we found that the TIGER and FTI posteriors are largely consistent across all deviation parameters, except for the 3PN logarithmic term. We also highlight results for the GW191204\_171526 event, which exhibited the largest inconsistency among the GWTC-3 events, with discrepancies visible in multiple PN terms. Since the baseline models of these two approaches differ, full agreement between the results is not expected. Additionally, the absence of higher-order modes and precession effects in the FTI model could contribute to the inconsistencies if the observed signal contains these components. 

%We note that the FTI analyses with the GW190814 event were performed using the \seobnrHMrom model to avoid the biases associated with neglecting the higher harmonics. The comparison against the TIGER is shown in the bottom row of Fig.~\ref{fig:tiger_vs_fti}. The shape of the posteriors for both cases is consistent for lower PN terms (-1PN to 1PN). The significant difference is evident for the higher PN terms starting from 2.5PN. For 3PN and 3.5PN, the GR value for the posterior from TIGER analysis is found outside the 90\% credible interval, whereas the FTI posteriors are consistent with GR.

We note that the FTI analyses of the GW190814 event were carried out using the \seobnrHMrom model~\cite{Cotesta:2020qhw} to mitigate biases arising from the omission of higher-order harmonics~\cite{LIGOScientific:2020tif}. A comparison with the TIGER results is presented in the bottom row of Fig.~\ref{fig:tiger_vs_fti}. The posterior distributions for both methods are consistent at lower PN orders (from -1PN to 1PN).  However, significant differences appear at higher PN orders, starting from 2.5PN. Notably, for the 3PN and 3.5PN terms, the GR value lies outside the 90\% credible interval in the TIGER analysis, while the FTI posteriors remain consistent with the GR predictions. For all three events, the differences are mostly noticeable for the higher PN terms. Since the baseline model of the FTI framework does not account for precession effects, their omission may potentially lead to inconsistencies.

%As shown in the bottom rows for the GW190814 event, the TIGER posterior is more consistent with the GR value.
% A noticeable change is seen for the -1PN parameter, where the FTI posterior peak is shifted away from the GR value, and its width is significantly wider than that of TIGER. We have also observed a similar difference for the GW190412 event. The strong contribution of higher-order modes in these two events leads to the improvement in TIGER posterior. Consequently, the combined posterior for the TIGER -1PN parameter, shown in Fig.~\ref{fig:violin_gwtc3_insp}, is significantly narrower than the FTI results reported in ~\cite{LIGOScientific:2021sio}.

\clearpage
\phantomsection

\bibliography{reference}
\end{document}